\documentclass[11pt,letterpaper]{article}
\usepackage[margin=1in]{geometry}
\usepackage{amsmath,amssymb,amsthm,mathtools}
\usepackage{lmodern}
\usepackage{microtype}
\usepackage{xcolor}
\usepackage{booktabs,tabularx,array}
\usepackage[colorlinks=true,linkcolor=primary,citecolor=accent,urlcolor=primary,bookmarksnumbered]{hyperref}
\expandafter\def\expandafter\UrlBreaks\expandafter{\UrlBreaks\do-\do\/\do.}

\definecolor{primary}{RGB}{24,52,92}
\definecolor{accent}{RGB}{15,105,100}

\hypersetup{
  pdftitle={Separating AC0 from Constant-Depth Quantum Circuits with Quantum Advice},
  pdfauthor={Xiaoming Sun, Jingheng Wang},
  pdfsubject={AC0 is not contained in BQAC0/qpoly}
}

\allowdisplaybreaks[2]

\theoremstyle{plain}
\newtheorem{theorem}{Theorem}[section]
\newtheorem{lemma}[theorem]{Lemma}
\newtheorem{corollary}[theorem]{Corollary}

\theoremstyle{definition}
\newtheorem{definition}[theorem]{Definition}

\theoremstyle{remark}
\newtheorem{remark}[theorem]{Remark}

\newcommand{\AC}{\mathsf{AC}^{0}}
\newcommand{\QAC}{\mathsf{QAC}^{0}}
\newcommand{\QACf}{\mathsf{QAC}^{0}_{f}}
\newcommand{\BQ}{\mathsf{BQAC}^{0}}
\newcommand{\EQ}{\mathsf{EQAC}^{0}}
\newcommand{\qpoly}{\mathsf{qpoly}}
\newcommand{\QAdv}{\BQ/\qpoly}
\newcommand{\EQAdv}{\EQ/\qpoly}
\newcommand{\PClass}{\mathsf{P}}
\newcommand{\ad}{\dagger}
\newcommand{\ket}[1]{\lvert #1\rangle}
\newcommand{\bra}[1]{\langle #1\rvert}
\newcommand{\braket}[2]{\langle #1\vert #2\rangle}
\newcommand{\ade}{\widetilde{\deg}}
\newcommand{\tr}{\operatorname{Tr}}
\newcommand{\supp}{\operatorname{supp}}
\newcommand{\wt}{\operatorname{wt}}
\newcommand{\poly}{\operatorname{poly}}
\newcommand{\norm}[1]{\left\lVert #1\right\rVert}
\newcommand{\wcoef}[1]{\mu_{N}(#1)}
\newcommand{\Pauli}{\mathcal{P}_{m}}
\newcommand{\hsigma}{\widehat{\sigma}}
\newcommand{\PhiZero}{\Phi_{0}}
\newcommand{\PhiHigh}{\Phi_{\ge R}}
\newcommand{\Pout}{P_{\mathrm{out}}}
\newcommand{\Qflt}{Q_{\mathrm{flt}}}
\newcommand{\PiPerp}{\Pi^{\perp}}
\newcommand{\cL}{\mathcal{L}}
\newcommand{\cI}{\mathcal{I}}
\newcommand{\bhi}{\beta_{\mathrm{hi}}}
\newcommand{\sloc}{\lambda_{\mathrm{loc}}}
\newcommand{\kw}{\kappa_{\mathrm{w}}}
\newcommand{\tcut}{\theta_{\mathrm{cut}}}
\newcommand{\iu}{\mathrm{i}}
\newcommand{\Ham}{\mathsf{H}}

\begin{document}

\title{\bfseries $\AC \not\subseteq \QAdv$ via Sublinear-Degree Pauli Approximation}
\author{%
  Xiaoming Sun$^{1,2}$ \qquad Jingheng Wang$^{1,2}$\\[0.6em]
  {\small $^{1}$State Key Lab of Processors, Institute of Computing Technology,}\\
  {\small Chinese Academy of Sciences, 100190, Beijing, China.}\\
  {\small $^{2}$School of Computer Science and Technology,}\\
  {\small University of Chinese Academy of Sciences, Beijing, 100049, China.}%
  \\[0.35em]
  {\small \texttt{sunxiaoming@ict.ac.cn} \quad
  \texttt{wangjinghang25s@ict.ac.cn}}%
}
\date{}
\maketitle

\begin{abstract}
We prove $\mathsf{AC}^0 \not\subseteq \mathsf{BQAC}^0/\mathsf{qpoly}$ in the single-copy, single-output model via an explicit $\mathsf{P}$-uniform family of linear-size monotone depth-2 CNFs. Every $\mathsf{QAC}^0$ circuit with arbitrary quantum advice achieves only an inverse-polynomial worst-case advantage over random guessing on this family. To prove this, we approximate the acceptance operator of any such circuit in operator norm by an advice-independent Hermitian operator of sublinear Pauli degree on a designated register alone, with degree exponent independent of width. Product-projection localization and Pauli twirling produce an auxiliary-register spectral cutoff that folds into contractions during backward peeling.

The same approximation yields two stretched-exponential bounds against $\mathsf{QAC}^0$ circuits on $m$-qubit states. First, given polynomially many copies and input-independent quantum advice, single-output circuits distinguish exact or exponentially approximate state $2$-designs from Haar-random states with advantage at most $e^{-m^\alpha}$. Second, any state prepared from fewer than $k$ arbitrary input qubits and remaining $|0\rangle$ inputs has spectral weight at most $2e^{-m^\alpha}$ in a constant-energy-density band of an Anshu--Breuckmann--Nirkhe NLTS Hamiltonian encoding $k$ logical qubits. Both bounds hold for every fixed $\alpha\in(0,1)$.
\end{abstract}
% \begin{abstract}
% We prove $\AC \not\subseteq \QAdv$ in the single-copy, single-output model via an explicit $\PClass$-uniform family of linear-size monotone depth-$2$ CNFs. Every constant-depth, polynomial-width $\QAC$ circuit with arbitrary quantum advice achieves only an inverse-polynomial worst-case advantage over random guessing on this family, with the exponent depending only on the depth.

% Our proof approximates the acceptance operator of any constant-depth, polynomial-width $\QAC$ circuit in operator norm by an advice-independent Hermitian operator of sublinear Pauli degree on a designated register alone, with exponent independent of width. Product-projection localization and Pauli twirling produce an auxiliary-register spectral cutoff that folds into contractions during backward peeling.

% The same approximation yields two $e^{-m^\alpha}$ bounds (for any fixed $\alpha < 1$) against such circuits on $m$-qubit states. Given polynomially many copies and input-independent quantum advice, single-output circuits distinguish exact or exponentially approximate state $2$-designs from Haar-random states with advantage at most $e^{-m^\alpha}$. Any state prepared from fewer than $k$ arbitrary input qubits and remaining $\ket{0}$ inputs has spectral weight at most $2e^{-m^\alpha}$ in a constant-energy-density band of an Anshu--Breuckmann--Nirkhe NLTS Hamiltonian encoding $k$ logical qubits.
% \end{abstract}

\section{Introduction}\label{sec:intro}

\subsection{Background and Motivation}

Moore~\cite{Moo99} introduced $\QAC$ as the quantum analogue of $\AC$: constant-depth, polynomial-size circuits built from arbitrary single-qubit unitaries and unbounded-fan-in generalized Toffoli gates (equivalently, multi-qubit controlled-$Z$ gates), without primitive unbounded fanout. While classical fanout allows an input bit to feed arbitrarily many clauses of a CNF, in constant-depth quantum circuits coherent fanout and $\mathrm{PARITY}$ are interreducible~\cite{Moo99}. Adjoining unbounded fanout collapses the shallow quantum hierarchy to $\QACf$, which contains $\AC$~\cite{HS05,TT16}.

Prior unconditional lower bounds against $\QAC$ follow two paradigms. In the Pauli-spectrum approach~\cite{NPVY24,ADOY25,DOY25}, operator-norm approximations of degree $\widetilde{O}(N^{1-3^{-d}})$~\cite{ADOY25} and $\widetilde{O}(N^{1-2^{-d}})$~\cite{DOY25} in the total circuit width $N$ guarantee sublinear degree in $n$ when $N = \widetilde{o}(n^{1/(1-2^{-d})})$. In a different direction, OpenAI~\cite{OAI26} introduced \emph{product-projection localization} (simplified in~\cite{GJL26,Kin26}) to prove $\mathrm{PARITY} \notin \BQ$. The same framework yields uniform-input Fourier concentration (both extending to $\QAdv$; see Remark~\ref{rem:fourier-advice}), as well as $\mathrm{MAJ}$ and symmetric-function lower bounds with $\ket{0}$ ancillas~\cite{OAI26,XL26}. Both $\mathrm{PARITY}$ and $\mathrm{MAJ}$, however, lie outside $\AC$~\cite{Ajt83,FSS84,Yao85,Has86}.

To separate $\AC$ from $\QAdv$ via Sherstov's near-linear approximate degree lower bound~\cite{She25}, we need a pointwise polynomial approximation of sublinear degree in the $m$ free variables that remain after restricting a universal monotone CNF inside a circuit of width $N = m^{O(1)}$. Uniform-input Fourier concentration~\cite{OAI26} does not give pointwise bounds, while total-width Pauli-degree bounds~\cite{ADOY25,DOY25} exceed $m$ at large polynomial width. We instead construct an \emph{asymmetric} operator-norm approximation: we bound the Pauli degree only on the $m$-qubit free-input register $X$ and absorb the localization cutoffs into contractions on the $(N-m)$-qubit auxiliary register $A$. Evaluating this advice-independent approximant on classical inputs gives the required polynomial.

In the reverse direction, for each fixed prime $p$, Grier, Morris, and Wu~\cite[Theorem~1]{GMW26} proved $\BQ \not\subseteq \AC[p]$ (where $\AC[p]$ adjoins $\mathrm{MOD}_p$ gates to $\AC$) for total Boolean functions in the single-copy decision model. Since $\AC \subseteq \AC[p]$ and $\BQ \subseteq \QAdv$, combining their result with $\AC \not\subseteq \QAdv$ shows that $\AC$ is incomparable with both $\BQ$ and $\QAdv$.

\subsection{Our Results}

\begin{theorem}[Main Separation]\label{th:separation}
There is an explicit $\PClass$-uniform family of monotone depth-$2$ CNFs $H = (H_n)_{n \ge 1}$, each with at most $3n$ literals and all $n$ variables relevant. For each fixed depth $d \ge 1$ there is a constant $\gamma_d > 0$ such that, for every fixed $c_0 \ge 1$ and all sufficiently large $n$, every $\QAC$ circuit of depth at most $d$ and width at most $(n+1)^{c_0}$ with arbitrary quantum advice has worst-case success probability at most $1/2 + O_d(n^{-\gamma_d})$ on $H_n$. In particular,
\[
  \AC \not\subseteq \QAdv.
\]
\end{theorem}

By contrast, diagonal (randomized classical) advice achieves $\Omega(1/n)$ worst-case advantage on $H_n$ in constant depth and width $O(n)$. For any $M$-clause CNF, let the advice sample a uniform clause, encoded by literal-indicator bits $(a_i^+, a_i^-)_{i=1}^n$, together with an independent bit $C$ with $\Pr[C=1] = a = 1/2 + 1/(4M)$. After one CNOT layer copies each input bit $x_i$, a layer of disjoint Toffoli gates writes $a_i^+ x_i$ and $a_i^-(1 - x_i)$ onto fresh ancillas; a generalized Toffoli controlled on their complements flips a $\ket{1}$ target iff all $2n$ products vanish (meaning the sampled clause is falsified), and a final Toffoli outputs the conjunction of this target with $C$. Every $1$-input satisfies all clauses and is accepted with probability $a$, whereas every $0$-input falsifies at least one clause and is rejected with probability at least $1 - a(1 - 1/M) = a + 1/(4M^2) > a$. Since $H_n$ has at most $3n$ clauses, this achieves worst-case advantage at least $1/(12n)$.

Theorem~\ref{th:separation} follows from the operator-norm approximation below.

\begin{theorem}[Advice-Independent Operator-Norm Approximation]\label{th:approximation}
Fix $d \in \mathbb{Z}_{\ge 1}$, $\kw \ge 1$, $\alpha \in (0,1)$, and $\eta > 0$. For all sufficiently large $m \ge m_0(d,\kw,\alpha,\eta)$ and every $N$-qubit $\QAC$ circuit $U$ of depth at most $d$ with $m \le N \le m^{\kw}$, partitioned into an $m$-qubit designated register $X$ and an $(N-m)$-qubit auxiliary register $A$, the acceptance operator $E = U^\ad \Pout U$ admits a Hermitian operator $B$ on $X \otimes A$ satisfying
\begin{equation}
  \norm{E - B} \le e^{-m^\alpha}
  \qquad\text{and}\qquad
  \deg_X(B) \le m^{\,1 - (1-\alpha)2^{-d} + \eta}.
  \label{eq:degree-bound}
\end{equation}
\end{theorem}

For a circuit computing $f$ with worst-case error at most $1/3$ and auxiliary state $\rho$, evaluating $B$ on $\ket{x}\bra{x} \otimes \rho$ yields a real polynomial of degree at most $\deg_X(B)$ approximating $f$ within $1/3 + e^{-m^\alpha}$. Rescaling and composing with the cubic majority polynomial restores error $1/3$.

\begin{corollary}[Approximate Degree Upper Bound for $\QAdv$]\label{cor:degree}
Fix $d \in \mathbb{Z}_{\ge 1}$, $\kw \ge 1$, and $\gamma > 0$, and set $\alpha = \min\{1/2, 2^{d-1}\gamma\}$. For all sufficiently large $m \ge m_1(d,\kw,\gamma)$, if $f \colon \{0,1\}^m \to \{0,1\}$ is computed with worst-case error at most $1/3$ by an $N$-qubit $\QAC$ circuit of depth at most $d$ with $m \le N \le m^{\kw}$ and arbitrary auxiliary state $\rho$ on $A$, then
\begin{equation}
  \ade_{1/3 + e^{-m^{\alpha}}}(f) \le m^{\,1 - 2^{-d} + \gamma}
  \qquad\text{and}\qquad
  \ade_{1/3}(f) \le 3 m^{\,1 - 2^{-d} + \gamma}.
  \label{eq:cor-degree}
\end{equation}
\end{corollary}

\paragraph{Pseudorandom states and stabilizer testing.}
Ghosh, Subramanian, and Zhan~\cite[Theorem~20 and Section~5]{GSZ26} proved that state $2$-designs are pseudorandom against $\mathsf{QNC}^0$ and conjectured security against $\QAC$. Theorem~\ref{th:state-designs} establishes the single-output case for exact and $2^{-\Omega(m)}$-approximate designs: given polynomially many copies of a sampled $m$-qubit state and arbitrary input-independent quantum advice, any single-output $\QAC$ circuit distinguishes such a design from Haar-random states with advantage at most $e^{-m^\alpha}$ for any fixed $\alpha \in (0,1)$. Corollary~\ref{cor:state-designs-multi} extends this bound in total variation distance to computational-basis readout on at most $m^\beta$ qubits for any fixed $\beta \in (0,1)$. For the random stabilizer and $T$-doped stabilizer ensembles of Section~\ref{sec:state-designs}, every state in the latter lies at trace distance at least $\sin(\pi/8)$ from all pure stabilizer states and the six-copy stabilizer test of~\cite{GNW21} separates them by a constant acceptance gap, yet single-output $\QAC$ circuits achieve advantage at most $2e^{-m^\alpha}$.

\paragraph{Low-energy state preparation.}
Theorem~\ref{th:low-energy-qac} extends the state-synthesis lower bound of~\cite[Corollary~6.8]{ADOY25} to arbitrary polynomial width and bounds spectral weight across the entire low-energy band. For an Anshu--Breuckmann--Nirkhe (ABN) NLTS Hamiltonian~\cite{ABN23} of a constant-rate, linear-distance CSS code encoding $k_m$ logical qubits, there is a constant energy density $e_0 > 0$ such that any state prepared by a $\QAC$ circuit from fewer than $k_m$ arbitrary input qubits and all remaining inputs in $\ket{0}$ has spectral weight at most $2e^{-m^\alpha}$ below energy $e_0 m$ (for any fixed $\alpha \in (0,1)$).

\subsection{Technical Overview}\label{sec:overview}

\paragraph{Auxiliary spectral cutoff via Pauli twirling.}
To prove Theorem~\ref{th:approximation}, we peel the $d$ Toffoli layers backward from $E = U^\ad \Pout U$. Conjugating a product projection $K = \bigotimes_{i=1}^N K_i$ by a single layer can inflate its support up to $N$. To compress these supports back down, we apply an integer-spectrum Chebyshev filter $q(D)$ to the mismatch counter $D = \sum_{i=1}^N (I - K_i)$, paired with an auxiliary projector $\Pi = I_X \otimes \pi$ acting solely on $A$ to suppress the high-mismatch region $[D \ge R]$ where $q(D)$ may grow large.

The central parameter trade-off is choosing the mismatch cutoff slightly superlinear in $m$, namely $R = \lceil m^{1+\tau} \rceil$ for a small $\tau > 0$. For a subcircuit $V$ of depth at most $d$, let $\PhiZero = V^\ad [D=0] V$ and $\PhiHigh = V^\ad [D \ge R] V$. Because $R$ exceeds $m$, padding to width $N_* = m^C$ and applying product-projection localization (Theorem~\ref{th:localization}) to each of the at most $2^m$ product projections in a Pauli string $\sigma \in \Pauli$ yields super-exponential decay $\norm{\PhiHigh (\sigma \otimes I_A) \PhiZero} \le 2^m e^{-m^\zeta}$ for $1 < \zeta < 1+\tau$ (Lemma~\ref{lem:pauli-transition}). Twirling $\PhiZero$ over $\Pauli$ removes all $X$-dependence and gives $\mathcal{T} = I_X \otimes S_A$ with $S_A = 2^{-m}\tr_X(\PhiZero)$; this satisfies $\PhiZero \preceq 4^m \mathcal{T}$ and $\PhiHigh \mathcal{T} \PhiHigh \preceq (2^m e^{-m^\zeta})^2 \PhiHigh$. Because $\zeta > 1$, the localization decay $e^{-m^\zeta}$ absorbs the $4^m$ twirling overhead: truncating $S_A$ at $e^{-m^\zeta}$ via $\pi = [S_A \ge e^{-m^\zeta}]$ gives $\Pi = I_X \otimes \pi$ with
\[
  \max\bigl\{\norm{(I-\Pi)\PhiZero},\, \norm{\PhiHigh \Pi}\bigr\} \le 2^m e^{-m^\zeta/2} \le e^{-m^{\sloc}}
\]
for $1 < \sloc < \zeta$ (Lemma~\ref{lem:selector}).

\paragraph{Backward peeling with a unified budget.}
Although $R = \lceil m^{1+\tau} \rceil$ is superlinear in $m$, both the Chebyshev filter degree and the logarithm of its $N$-weighted coefficient norm scale as the geometric mean $\widetilde{O}(\sqrt{R s_{j-1}})$ of $R$ and the current budget $s_{j-1}$. Factoring $U$ from output to input as $V_0 = U$ and $V_{j-1} = G_j V_j$ ($1 \le j \le d$), where $G_j$ is the $j$-th Toffoli layer, we maintain Hermitian approximants
\[
  E_j = \sum_\iota c_\iota (I_X \otimes C_\iota) V_j^\ad K_\iota V_j (I_X \otimes C_\iota^\ad),
  \qquad c_\iota \in \mathbb{R},\quad \norm{C_\iota} \le 1,
\]
controlled by a single budget $s_j$ bounding both $|\supp K_\iota| \le s_j$ and $\sum_\iota |c_\iota| \le e^{s_j}$, starting from $s_0 = m^\alpha + \log d$. At step $j$, expanding $G_j^\ad K_\iota G_j$ via Lemma~\ref{lem:layer} multiplies the $\ell_1$ weight by $18^{s_{j-1}}$ to at most $e^{4 s_{j-1}}$; setting the filter parameter $r_j = \lceil 3 s_{j-1} \rceil$ absorbs this growth, while $e^{-m^{\sloc}}$ suppresses the high-mismatch tails. Because each new projector $\Pi = I_X \otimes \pi$ acts only on $A$, it folds directly into the outer contraction via $C_\iota \leftarrow C_\iota \pi$ without affecting $X$ or increasing $\norm{C_\iota} \le 1$. Unrolling the recurrence gives
\[
  s_j = \widetilde{O}\bigl(\sqrt{R s_{j-1}}\bigr),
  \qquad
  s_d = \widetilde{O}\bigl(R^{1 - 2^{-d}} s_0^{2^{-d}}\bigr) = \widetilde{O}\bigl(m^{(1+\tau)(1-2^{-d}) + \alpha 2^{-d}}\bigr).
\]
Choosing $\tau > 0$ small relative to $(1-\alpha)2^{-d}$ keeps $s_j = o(m)$ at every step. At $j = d$, $V_d$ consists solely of single-qubit unitaries and the contractions $C_\iota$ live on $A$, so $\deg_X(E_d) \le s_d \le m^{1 - (1-\alpha)2^{-d} + \eta}$.

\paragraph{All-lengths monotone universal selector CNF.}
For every $\delta \in (0,1)$, Sherstov~\cite[Theorem~5.1]{She25} constructs monotone width-$w(\delta)$ CNFs $F_m$ with $\ade_{1/2 - m^{-2}}(F_m) = \Omega_\delta(m^{1-\delta})$. To accommodate the clause widths needed across all circuit depths, we allocate disjoint blocks of sizes $B_w = \lfloor n/2^w \rfloor$ ($w \ge 1$) at each input length $n$. In each active block, we place the largest monotone universal selector CNF $H_{m_w, w}$ that fits in $B_w$ bits and add unit clauses on all unused bits (Section~\ref{sec:universal-cnf}); the resulting formula $H_n$ has at most $3n$ literals. For any fixed $(d, c_0)$ and $\delta = 2^{-d}/3$, we have $m_w = \Theta_w(n^{1/w})$ for all sufficiently large $n$; restricting $H_n$ to block $w$ embeds $F_{m_w}$ into a circuit of width at most $m_w^{w c_0 + 1}$, for which Theorem~\ref{th:approximation} rules out worst-case success probability at least $1/2 + 2m_w^{-2} = 1/2 + O_d(n^{-2/w})$.

\paragraph{Pseudorandomness with multiple copies.}
By a hybrid argument, we fix the sampled state and replace its $t$ copies one at a time by the maximally mixed state. At each step, we place the remaining copies and advice in the auxiliary register and apply Theorem~\ref{th:approximation} to bound the Pauli degree on register $X$ holding the copy being replaced. By Cauchy--Schwarz, the change in acceptance probability is controlled by the operator-norm error and the square root of the sampled state's low-weight Pauli mass, which is small on average by the second-moment condition of a state $2$-design. Summing over all $t$ steps compares both the design and Haar ensembles to the same product of maximally mixed states.

\paragraph{Low-energy spectral weight.}
Inverting the preparation circuit and testing its $\ket{0}$ inputs with a flag qubit gives a single-output test for the $2^b$-dimensional preparation subspace. We approximate this test in operator norm with Pauli degree $o(m)$ on the $m$ designated output qubits. Using the ABN clustering property and linear code distance, we choose bases on the low-energy syndrome eigenspaces in which the approximant, compressed to the low-energy band, acts as the identity on the $k_m$ logical qubits. The resulting $2^{k_m}$-fold eigenvalue multiplicity, combined with $b < k_m$, bounds the overlap between the preparation subspace and the entire low-energy band.

\subsection{Related Work}

Table~\ref{tab:related} compares our separation with prior unconditional lower bounds against $\QAC$.

\begin{table}[t!]
\centering
\footnotesize
\setlength{\tabcolsep}{6pt}
\renewcommand{\arraystretch}{1.22}
\begin{tabularx}{\linewidth}{@{}>{\raggedright\arraybackslash}p{3.55cm} >{\raggedright\arraybackslash}p{4.65cm} >{\raggedright\arraybackslash}p{2.95cm} >{\raggedright\arraybackslash}X@{}}
\toprule
\textbf{Work} & \textbf{Target / Bound} & \textbf{Ancilla / Advice} & \textbf{Depth / Width Regime} \\
\midrule
Fang et~al.~\cite{FFGHZ06} & Exact clean $\mathrm{PARITY}$ & $\ket{0}$ ancillas & Const.\ $d$, $a = o(n)$ \\
\addlinespace[2pt]
Pad\'e et~al.~\cite{PFGT20}\newline Fenner et~al.~\cite{FGPT25} & Exact $\mathrm{PARITY}$ (clean~\cite{PFGT20} / unclean~\cite{FGPT25}) & $\ket{0}$ ancillas & $d = 2$, arbitrary $a$ \\
\addlinespace[2pt]
Rosenthal~\cite{Ros21} & Approx.\ $\mathrm{PARITY}$ (input-preserving) & $\ket{0}$ ancillas & $d = 2$, arbitrary $a$ \\
\addlinespace[2pt]
Nadimpalli et~al.~\cite{NPVY24} & $\mathrm{PARITY}$, $\mathrm{MAJ}$ & Input-independent state & Const.\ $d$, $a = o(n^{1/d})$ \\
\addlinespace[2pt]
Anshu et~al.~\cite{ADOY25} & Operator-norm approx.\ ($\widetilde{O}(N^{1-3^{-d}})$) & Input-independent state & Const.\ $d$, $N = \widetilde{o}(n^{1/(1-3^{-d})})$ \\
\addlinespace[2pt]
Dong et~al.~\cite{DOY25} & Operator-norm approx.\ ($\widetilde{O}(N^{1-2^{-d}})$) & Input-independent state & Const.\ $d$, $N = \widetilde{o}(n^{1/(1-2^{-d})})$ \\
\addlinespace[2pt]
Joshi et~al.~\cite{JTVW26} & Exact $\mathrm{PARITY}$ (any size), exact $\mathrm{MAJ}$ (gate lower bound);\newline $\mathrm{PARITY}$ corr.\ $e^{-\Omega(\sqrt{n})}$ & Separable / $\ket{0}$ ancillas & Exact $d = 3$;\newline corr.\ $d = 2$ \\
\addlinespace[2pt]
OpenAI~\cite{OAI26}; Gao et~al.~\cite{GJL26};\newline Kintali~\cite{Kin26} & $\mathrm{PARITY}$; Fourier tail~\cite[Cor.~9]{OAI26};\newline $\mathrm{MAJ}$/sym.~\cite[Cor.~10]{OAI26} (via~\cite{XL26});\newline avg.-case $\mathrm{MAJ}$~\cite[Cor.~11]{OAI26} & $\ket{0}$ ancillas\newline ($\mathrm{PARITY}$, Fourier tail: $\qpoly$) & Const.\ $d$, $\poly(n)$ width \\
\midrule
\textbf{This work} & \textbf{Monotone depth-$2$ CNF ($\AC$);}\newline \textbf{Pauli degree on $X$ (operator norm)} & \textbf{Arbitrary $\qpoly$ advice} & \textbf{Const.\ $d$, $\poly(n)$ width} \\
\bottomrule
\end{tabularx}
\caption{Comparison with prior unconditional lower bounds against $\QAC$ circuits ($n$ input bits, $a$ ancillas, total width $N$).}
\label{tab:related}
\end{table}

\paragraph{Shallow quantum advantage and $\BQ$ vs.\ $\AC[p]$.}
Early separations of shallow quantum circuits from $\mathsf{NC}^0$ and $\AC$ targeted search and relational problems~\cite{BGK18,LG19,BGKT20,WKST19}. For decision problems, Grier, Morris, and Wu~\cite[Theorems~1 and~2]{GMW26} proved $\BQ \not\subseteq \AC[p]$ for each fixed prime $p$ in the single-copy model, and showed that $\QAC$ circuits given $\poly(n)$ copies of the classical input compute $\mathsf{TC}^0$ with bounded error. On the upper-bound side for standard $\QAC$ (see also~\cite{GHMP02} on $\mathsf{QACC}^0$), Rosenthal~\cite{Ros21} constructed constant-depth, exponential-size approximations to parity and fanout, Foxman et~al.~\cite{FPVY26} gave constant-depth approximate unitary designs, and Gretta, Gupta, and Joshi~\cite{GGJ26} built $\QAC$ circuits achieving $1 - o(1)$ average-case correlation with $\mathrm{MAJ}$.

\paragraph{Pauli spectrum analysis and operator-norm approximations.}
Nadimpalli, Parham, Vasconcelos, and Yuen~\cite{NPVY24} initiated the Pauli-spectrum analysis of $\QAC$: they proved low-degree Pauli concentration in Frobenius norm for $a = o(n^{1/d})$ ancillas and deduced $\mathrm{PARITY}$ and $\mathrm{MAJ}$ lower bounds. Anshu, Dong, Ou, and Yao~\cite[Lemma~3.5]{ADOY25} and Dong, Ou, and Yao~\cite[Theorem~3.1]{DOY25} strengthened this framework to \emph{operator-norm} low-degree Pauli approximations of degree $\widetilde{O}(N^{1-3^{-d}})$ and $\widetilde{O}(N^{1-2^{-d}})$, respectively. Anshu et~al.~\cite[Corollary~6.8]{ADOY25} obtained an ancilla lower bound for low-energy state synthesis, and Dong et~al.~\cite{DOY25} gave learning and testing algorithms for linear-size $\QAC$ channels. In parallel, Vasconcelos and Huang~\cite{VH25} used low-degree Pauli structure to learn logarithmic-ancilla $\QAC$ circuits, and Gretta, Gupta, and Joshi~\cite{GGJ26} proved that $\mathrm{PARITY} \notin \BQ$ is equivalent to a quantum Linial--Mansour--Nisan (LMN) Fourier-concentration conjecture. Theorem~\ref{th:approximation} removes the total-width bottleneck of~\cite{ADOY25,DOY25} by restricting the Pauli degree truncation to the designated register $X$ at arbitrary polynomial width.

\paragraph{Product-projection localization.}
Product-projection localization~\cite{OAI26} (Theorem~\ref{th:localization}) bounds the off-diagonal block of $V[D=0]V^\ad$ by $\norm{[M \ge N^{\theta_2}] V [D=0] V^\ad [M=0]} \le e^{-N^{\theta_1}}$ for $\QAC$ circuits $V$ of depth at most $d$, with fixed $d$ and $0 \le \theta_1 < \theta_2$ and sufficiently large $N$. Gao, Ji, and Liu~\cite{GJL26} simplified the proof by direct layer removal, and Kintali~\cite{Kin26} gave a polarization-based proof (our real-coefficient single-layer expansion in Lemma~\ref{lem:layer} uses the same polarization identity). From localization,~\cite{OAI26} deduced $\mathrm{PARITY} \notin \BQ$ and uniform-input Fourier concentration~\cite[Corollary~9]{OAI26} (Remark~\ref{rem:fourier-advice} records the extension to $\QAdv$), which together with the Fourier tail of $\mathrm{MAJ}$ gives a uniform average-case lower bound~\cite[Corollary~11]{OAI26}. Combining the parity lower bound with the reductions of~\cite{XL26} further yields worst-case lower bounds for $\mathrm{MAJ}$ and symmetric functions~\cite[Corollary~10]{OAI26}. Our Pauli-twirling spectral cutoff (Section~\ref{sec:twirl}) lifts localization from uniform-input Fourier tails to advice-independent operator-norm approximations.

\paragraph{Approximate degree and quantum advice.}
Approximate degree~\cite{NS94} entered quantum lower bounds through the polynomial method~\cite{BBCMW01}, with tight small-error characterizations for symmetric functions such as $\ade_{e^{-m^\alpha}}(\mathrm{AND}_m) = \Theta(m^{(1+\alpha)/2})$~\cite{dW08}. For classical $\AC$, Bun and Thaler~\cite{BT17} proved an $\Omega(n^{1-\delta})$ approximate degree lower bound, and Sherstov~\cite{She25} achieved $\Omega_\delta(n^{1-\delta})$ already for depth-$2$ monotone constant-width DNFs/CNFs at error $1/2 - n^{-2}$. Conversely, for fixed clause width $w$, every width-$w$ CNF has constant-error approximate degree $O_w(n^{1-1/(w+1)})$~\cite{She20}. Both the constant clause width and the $1/2 - n^{-2}$ error regime enable our linear-size universal selector CNF and $1/2 + O_d(n^{-\gamma_d})$ separation. Quantum advice ($\qpoly$) was introduced by Nishimura and Yamakami~\cite{NY04} and studied extensively in~\cite{Aar05,AD14,ABK24}; in the shallow-circuit setting, Slote~\cite{Slo24} proved parity lower bounds for linear-size $\AC \circ \mathsf{QNC}^0/\qpoly$.

\paragraph{Organization.}
Section~\ref{sec:prelim} sets up the models and notation. Section~\ref{sec:approx} proves Theorem~\ref{th:approximation}. Section~\ref{sec:separation} proves Corollary~\ref{cor:degree}, constructs the monotone universal selector CNF, and deduces Theorem~\ref{th:separation}. Section~\ref{sec:quantum-applications} establishes the pseudorandomness of state $2$-designs and the low-energy state preparation bound. Section~\ref{sec:open} concludes with open problems.

\section{Preliminaries}\label{sec:prelim}

Write $[n] = \{1, \dots, n\}$. We use $\log$ for the natural logarithm and $\log_2$ for the base-$2$ logarithm. The notation $\norm{\cdot}$ denotes the Euclidean norm on vectors and the operator norm on operators. The trace norm is $\norm{Y}_1 = \tr\sqrt{Y^\ad Y}$. The trace distance between density operators $\rho$ and $\sigma$ is $\frac12\norm{\rho - \sigma}_1$. The total variation distance between discrete distributions $p$ and $q$ is $\frac12\sum_y |p(y) - q(y)|$.

For Hermitian operators $Y_1$ and $Y_2$, write $Y_1 \preceq Y_2$ if $Y_2 - Y_1$ is positive semidefinite. For a Hermitian operator $Y$ and $\cI \subseteq \mathbb{R}$, the spectral projector $[Y \in \cI]$ projects onto the eigenspaces of $Y$ with eigenvalues in $\cI$. We write, for example, $[Y = 0]$ and $[Y \ge \theta]$ for the projectors associated with $\{0\}$ and $[\theta,\infty)$.

\begin{definition}[$\AC$ and CNF Formulas]\label{def:ac0}
$\AC$ consists of Boolean function families computed by constant-depth, polynomial-size circuits over input literals and unbounded-fan-in AND and OR gates, with unbounded fanout.

A \emph{CNF formula} is a conjunction of clauses, each a disjunction of literals. It is \emph{monotone} if all literals are positive. Its \emph{width} is the maximum clause length, and its \emph{size} is the total number of literal occurrences. For $f \colon \{0,1\}^n \to \{0,1\}$, input bit $i$ is \emph{relevant} if there are two inputs that differ only at $i$ and on which $f$ takes different values. A formula family $(F_n)_{n \ge 1}$ is \emph{$\PClass$-uniform} if a deterministic algorithm outputs $F_n$ on input $1^n$ in time $\poly(n)$.
\end{definition}

\begin{definition}[$\QAC$ Circuits and $\QAdv$]\label{def:bqac0}
The \emph{width} $N$ of a $\QAC$ circuit is its total number of qubits. A circuit of \emph{depth} $d$ alternates $d+1$ layers of arbitrary single-qubit unitaries with $d$ layers of generalized Toffoli gates. Gates in a layer act on disjoint sets of qubits. A generalized Toffoli gate with $k \ge 1$ controls acts by
\[
  \ket{x,y} \longmapsto \ket{x,y \oplus \bigwedge_{i=1}^k x_i},
  \qquad x \in \{0,1\}^k,\quad y \in \{0,1\}.
\]
A Hadamard on the target before and after the gate gives a multi-qubit controlled-$\mathrm{Z}$ gate.

A $\QAC$ circuit family has constant depth and total width polynomial in the problem size. This size is the input length $n$ for Boolean functions and the number $m$ of qubits in the state being distinguished or prepared in Section~\ref{sec:quantum-applications}.

For a circuit $U$ with one designated output qubit, acceptance is the computational-basis outcome $1$ on that qubit. The output projector is $\Pout = \ket{1}\bra{1}_{\mathrm{out}} \otimes I$, and the \emph{acceptance operator} is $E = U^\ad \Pout U$. The acceptance probability on an $N$-qubit state $\sigma$ is $\tr(E\sigma)$.

In the single-copy, single-output decision model, the input $z \in \{0,1\}^n$ is encoded by one copy of $\ket{z}$. The circuit $U_n$ also receives an input-independent advice state $\varrho_n$ on $b_n$ qubits and $a_n$ ancillas initialized to $\ket{0}$, for a total width $N_n = n + b_n + a_n$. Its acceptance probability is
\[
  p_{\mathrm{acc}}(z) = \tr\bigl(E_n(\ket{z}\bra{z} \otimes \varrho_n \otimes \ket{0^{a_n}}\bra{0^{a_n}})\bigr),
  \qquad E_n = U_n^\ad \Pout U_n.
\]
For $f_n \colon \{0,1\}^n \to \{0,1\}$, the \emph{worst-case error} is $\varepsilon_n = \max_{z \in \{0,1\}^n} |p_{\mathrm{acc}}(z) - f_n(z)|$. The \emph{worst-case success probability} is $1-\varepsilon_n$, and the \emph{worst-case advantage} over random guessing is $1/2-\varepsilon_n$.

A family $f = (f_n)_{n \ge 1}$ lies in $\QAdv$ if there are constants $d, c_0 \ge 1$, circuits $(U_n)_{n \ge 1}$ of depth at most $d$ and width $N_n \le (n+1)^{c_0}$, and advice states $(\varrho_n)_{n \ge 1}$ with $\varepsilon_n \le 1/3$ for all $n \ge 1$. The exact class $\EQAdv$ requires $\varepsilon_n = 0$ for every $n$. With $b_n = 0$ for every $n$, the bounded-error and exact classes are $\BQ$ and $\EQ$, respectively.
\end{definition}

\begin{definition}[Approximate Degree]\label{def:approx-deg}
For $f \colon \{0,1\}^m \to \{0,1\}$ and $\varepsilon \in (0,1/2)$, the \emph{$\varepsilon$-approximate degree} $\ade_\varepsilon(f)$ is the minimum total degree of a real polynomial $p$ satisfying $|p(x) - f(x)| \le \varepsilon$ for every $x \in \{0,1\}^m$.
\end{definition}

\begin{definition}[Designated Register and Pauli Degree]\label{def:pauli-deg}
Partition the $N$ qubits into an $m$-qubit \emph{designated register} $X$ and an $(N-m)$-qubit \emph{auxiliary register} $A$. The Pauli strings on $X$ are $\Pauli = \{I,\mathrm{X},\mathrm{Y},\mathrm{Z}\}^{\otimes m}$, where $\mathrm{X},\mathrm{Y},\mathrm{Z}$ are the single-qubit Pauli matrices. For $\sigma \in \Pauli$, its weight $\wt(\sigma)$ is the number of nonidentity factors, and $\hsigma = \sigma \otimes I_A$ is its extension to $X \otimes A$.

The orthogonality relation $\tr(\sigma \sigma') = 2^m\delta_{\sigma,\sigma'}$ gives the unique expansion of an operator $B$ on $X \otimes A$,
\[
  B = \sum_{\sigma \in \Pauli} \sigma \otimes B_\sigma,
  \qquad B_\sigma = 2^{-m}\tr_X(\hsigma B).
\]
Here $\tr_X$ is the partial trace over $X$. Each coefficient $B_\sigma$ acts on $A$ and is Hermitian when $B$ is Hermitian. The \emph{Pauli degree on $X$} of a nonzero $B$ is
\[
  \deg_X(B) = \max\{\wt(\sigma) : \sigma \in \Pauli,\ B_\sigma \ne 0\}.
\]
Set $\deg_X(0) = 0$.
\end{definition}

\begin{definition}[Product Projections and Mismatch Counters]\label{def:mismatch}
A \emph{product projection} on $N$ qubits is an orthogonal projector $K = \bigotimes_{i=1}^N K_i$, where each $K_i$ is the identity or a single-qubit rank-$1$ projector. Its \emph{support} is $\supp K = \{i \in [N] : K_i \ne I\}$. The associated \emph{mismatch counter} is
\[
  D = \sum_{i=1}^N (I - K_i) = \sum_{i \in \supp K} (I - K_i),
\]
where each summand acts as the identity on the other qubits. On a tensor product of eigenvectors of the $K_i$, the eigenvalue of $D$ counts the positions at which the local projection has eigenvalue zero. Thus $\operatorname{spec}(D) \subseteq \{0, \dots, |\supp K|\} \subseteq \{0, \dots, N\}$ and $K = [D = 0]$.
\end{definition}

\section{Advice-Independent Operator-Norm Approximation}\label{sec:approx}

To prove Theorem~\ref{th:approximation}, we collect the preliminary tools in Section~\ref{sec:basis}, construct the auxiliary separating projector via Pauli twirling in Section~\ref{sec:twirl}, establish the low-support replacement in Section~\ref{sec:replacement}, and run the backward layer induction in Section~\ref{sec:induction}.

\subsection{Basic Tools}\label{sec:basis}

\begin{theorem}[Product-Projection Localization~{\cite[Theorem~2]{OAI26}}; see also~\cite{GJL26,Kin26}]\label{th:localization}
Fix $d \in \mathbb{Z}_{\ge 0}$ and $0 \le \theta_1 < \theta_2$. For all sufficiently large $N \ge N_0(d, \theta_1, \theta_2)$, every $N$-qubit $\QAC$ circuit $V$ of depth at most $d$, and every pair of mismatch counters $M, D$ on $N$ qubits,
\begin{equation}
  \norm{[M \ge N^{\theta_2}]\, V [D = 0] V^\ad [M = 0]} \le e^{-N^{\theta_1}}.
  \label{eq:localization}
\end{equation}
\end{theorem}

\begin{remark}[Fourier Concentration and Parity with Quantum Advice]\label{rem:fourier-advice}
Theorem~\ref{th:localization} directly extends~\cite[Corollary~9]{OAI26} to arbitrary fixed quantum advice. For a pure state $\ket{a}$ on the $(N-n)$-qubit auxiliary register, $p(x) = \bra{x,a}E\ket{x,a}$ ($E = U^\ad \Pout U$), and Hadamard basis states $\ket{\xi_h} = 2^{-n/2}\sum_{x\in\{0,1\}^n}(-1)^{h\cdot x}\ket{x}$ ($h \in \{0,1\}^n$), let $\mathbf{1}_S \in \{0,1\}^n$ denote the indicator string of $S \subseteq [n]$. Expanding in the computational basis and using $\mathbb{E}_h[(-1)^{h\cdot(u+x)}] = \delta_{u,x}$ for uniform $h$ gives $\widehat{p}(S) = 2^{-n}\sum_x (-1)^{\mathbf{1}_S\cdot x}p(x) = \mathbb{E}_h\bra{\xi_{h\oplus\mathbf{1}_S},a}E\ket{\xi_h,a}$. Letting $M_h$ count mismatches from the factors of $\ket{\xi_h}$ on the $n$ input qubits (acting as $I$ on the auxiliary register), $\ket{\xi_h,a}$ lies in $\operatorname{im}[M_h=0]$, while $\{\ket{\xi_{h\oplus\mathbf{1}_S},a} : |S| \ge k\}$ is an orthonormal set in $\operatorname{im}[M_h \ge k]$. By Jensen's inequality,
\[
  \sum_{|S|\ge k}|\widehat{p}(S)|^2
  \le \mathbb{E}_h\sum_{|S|\ge k}\bigl|\bra{\xi_{h\oplus\mathbf{1}_S},a}E\ket{\xi_h,a}\bigr|^2
  \le \mathbb{E}_h\norm{[M_h\ge k]E[M_h=0]}^2,
\]
and convexity of $|\widehat{p}(S)|^2$ extends this bound to any mixed auxiliary state. If $n \le N \le (n+1)^{c_0}$ and $k = \lceil n^\delta \rceil$ ($\delta \in (0,1]$), choosing $0 < \theta_1 < \theta_2 < \delta/c_0$ gives $N^{\theta_2} \le k$ for all sufficiently large $n$, so~\eqref{eq:localization} with $V = U^\ad$ and $[D=0] = \Pout$ bounds the tail weight by $e^{-2N^{\theta_1}}$. In particular, worst-case advantage $\varepsilon > 0$ on $\mathrm{PARITY}_n$ would give $\mathbb{E}_x[(-1)^{\mathrm{PARITY}_n(x)}(1 - 2p(x))] = -2\widehat{p}([n]) \ge 2\varepsilon$, yielding $\mathrm{PARITY} \notin \QAdv$.
\end{remark}

We also need two elementary lemmas: a single-layer expansion for generalized Toffoli gates and a Chebyshev filter for mismatch counters.

\begin{lemma}[Real-Coefficient Single-Layer Expansion; cf.~{\cite[Lemma~3]{GJL26}}, {\cite[Lemma~3]{Kin26}}]\label{lem:layer}
Let $K$ be an $N$-qubit product projection with $|\supp K| \le k$, and let $G$ be a single layer of pairwise disjoint generalized Toffoli gates preceded and followed by layers of arbitrary single-qubit unitaries. Then $G^\ad K G$ is a \emph{real} linear combination of product projections with $\ell_1$ weight at most $18^k$.
\end{lemma}

\begin{proof}
Single-qubit layers preserve product projections and their support sizes under conjugation, so it suffices to analyze a layer of disjoint generalized Toffoli gates.

Since $\mathrm{X} = I - 2\ket{-}\bra{-}$, each gate $g$ in $G$ takes the form $g = I - 2\Pi_g$, where $\Pi_g$ acts as $\ket{1}\bra{1}$ on the controls of $g$, $\ket{-}\bra{-}$ on its target, and $I$ elsewhere. Gates disjoint from $\supp K$ commute with $K$ and with all other gates in $G$, so they cancel in $G^\ad K G$. Let $\mathcal{G}$ be the remaining set of gates; pairwise disjointness gives $|\mathcal{G}| \le k$. Writing $\Pi_S = \prod_{g\in S}\Pi_g$, expanding both sides of $K$ yields
\[
  G^\ad K G
  = \sum_{S,T\subseteq\mathcal{G}}(-2)^{|S|+|T|}\Pi_S K\Pi_T,
  \qquad
  \sum_{S,T\subseteq\mathcal{G}}2^{|S|+|T|}=3^{2|\mathcal{G}|}\le 9^k.
\]

Fix any nonzero term $\Pi_S K \Pi_T$, whose $i$-th tensor factor is $P_i K_i P'_i$ with $P_i, P'_i \in \{I, (\Pi_g)_i\}$ if some $g \in \mathcal{G}$ acts on qubit $i$ and $P_i = P'_i = I$ otherwise. For $i \notin \supp K$, $K_i = I$ and idempotence leaves $I$ or a rank-$1$ projector. For $i \in \supp K$ with $K_i = \ket{w}\bra{w}$, the nonzero rank-$1$ factor $P_i K_i P'_i = (P_i\ket{w})(P'_i\ket{w})^\ad$ has norm $\lambda \in (0, 1]$ and thus equals $\lambda\ket{u}\bra{v}$ for unit vectors $\ket{u}, \ket{v}$. Since $\sum_{j=0}^3 \iu^j = \sum_{j=0}^3 \iu^{2j} = 0$, the four-phase polarization identity gives
\[
  \ket{u}\bra{v} = \frac{1}{4}\sum_{j=0}^{3} \iu^j \ket{u + \iu^j v}\bra{u + \iu^j v}.
\]
Normalizing each nonzero vector $\ket{u + \iu^j v}$ into a rank-$1$ projector contributes absolute weight $\frac{\lambda}{4}\norm{\ket{u} + \iu^j\ket{v}}^2$, summing to
\[
  \frac{\lambda}{4}\sum_{j=0}^{3} \norm{\ket{u} + \iu^j \ket{v}}^2
  = \frac{\lambda}{4}\sum_{j=0}^{3}\bigl(2 + 2\operatorname{Re}(\iu^j\braket{u}{v})\bigr)
  = 2\lambda \le 2.
\]
Applying this on each of the at most $k$ qubits in $\supp K$ multiplies the $\ell_1$ weight by at most $2^k$, giving $G^\ad K G = \sum_\iota \tilde{c}_\iota P_\iota$ with $\sum_\iota |\tilde{c}_\iota| \le 9^k \cdot 2^k = 18^k$. Since $G^\ad K G$ and every $P_\iota$ are Hermitian, taking the Hermitian part gives $G^\ad K G = \sum_\iota (\operatorname{Re}\tilde{c}_\iota) P_\iota$ with $\sum_\iota |\operatorname{Re}\tilde{c}_\iota| \le 18^k$.
\end{proof}

For a real polynomial $q(x)=\sum_j q_j x^j$, define its \emph{$N$-weighted coefficient norm} $\wcoef{q}=\sum_j|q_j|N^j$. By Cauchy product of coefficients, $\wcoef{pq} \le \sum_{i,j}|p_i q_j|N^{i+j} = \wcoef{p}\wcoef{q}$, and for any $N$-qubit mismatch counter $D$, $\norm{D} \le N$ implies $\norm{q(D)} \le \sum_j |q_j|\norm{D}^j \le \wcoef{q}$.

\begin{lemma}[Chebyshev Integer-Spectrum Filter; cf.~{\cite[Lemma~3]{OAI26}}, {\cite[Lemma~5]{GJL26}}]\label{lem:polynomial}
Let $N, R, r$ be positive integers satisfying $1 \le r \le R/2$, and set $\Lambda_0 = 2 + N + R$. There exists a real polynomial $q$ such that
\begin{align}
  q(0) &= 1, \qquad |q(j)| \le e^{-3r} \quad (j \in \{1, \dots, R\}), \label{eq:filter-error}\\
  \deg q &\le 11\sqrt{Rr}\log \Lambda_0, \qquad
  \log \wcoef{q} \le 11\sqrt{Rr}\log^2 \Lambda_0. \label{eq:filter-weight}
\end{align}
Moreover, for any $N$-qubit mismatch counter $D$, the operator $q(D)$ expands as a real linear combination of product projections of support at most $\deg q$ with total $\ell_1$ weight at most $\wcoef{q}$.
\end{lemma}

\begin{proof}
Let $T_h$ be the degree-$h$ Chebyshev polynomial of the first kind, defined by $T_0(z) = 1$, $T_1(z) = z$, and $T_{h+1}(z) = 2z T_h(z) - T_{h-1}(z)$, so that $T_h(\cos u) = \cos(hu)$ and $T_h(\cosh u) = \cosh(hu)$. Set $h = \lceil 4\sqrt{Rr}\log \Lambda_0 \rceil$ and $\varphi(x) = (R + r - 2x)/(R - r)$, which maps $[r, R]$ bijectively onto $[-1, 1]$, and define
\[
  q(x) = \prod_{j=1}^{r}\!\left(1 - \frac{x}{j}\right) \cdot \frac{T_h(\varphi(x))}{T_h(\varphi(0))}.
\]
Clearly $q(0) = 1$ and $q(j) = 0$ for all integers $1 \le j \le r$. For $x \in (r, R]$, we have $\varphi(x) \in [-1, 1]$, hence $|T_h(\varphi(x))| \le 1$, while $\bigl|\prod_{j=1}^r(1 - x/j)\bigr| \le (1 + R)^r$. Setting $\xi = \sqrt{r/R} \in (0, 1/\sqrt{2}]$, we obtain
\[
  \varphi(0) = \frac{1 + \xi^2}{1 - \xi^2} = \cosh\!\left(\log\frac{1+\xi}{1-\xi}\right),
  \qquad \log\frac{1+\xi}{1-\xi} = \int_0^\xi \frac{2}{1 - u^2}\,\mathrm{d}u \ge 2\xi,
\]
so $T_h(\varphi(0)) = \cosh\bigl(h\log\frac{1+\xi}{1-\xi}\bigr) \ge \frac{1}{2}e^{2h\sqrt{r/R}}$. Since $1 \le r \le R/2$ and $N \ge 1$ imply $R \ge 2$ and $\Lambda_0 = 2 + N + R \ge 5$, it follows that
\[
  |q(x)| \le 2(1 + R)^r e^{-2h\sqrt{r/R}} \le 2 \Lambda_0^{r - 8r} = 2\Lambda_0^{-7r} \le e^{-3r},
\]
since $2 \cdot 5^{-7} < e^{-3}$, which proves~\eqref{eq:filter-error}.

Since $\sqrt{Rr}\log\Lambda_0\ge\sqrt{2}\log 5>1$, we have $h\le 5\sqrt{Rr}\log\Lambda_0$ and $r \le \sqrt{Rr}$, so $\deg q = r + h \le 6\sqrt{Rr}\log\Lambda_0 \le 11\sqrt{Rr}\log\Lambda_0$. Because $R-r\ge R/2\ge 1$, the constant and linear coefficients of $\varphi$ have magnitudes at most $3$ and $2$, giving $\wcoef{\varphi}\le 3+2N$. Set $\Lambda_\varphi=1+2\wcoef{\varphi}\le 4\Lambda_0$, so $\log\Lambda_\varphi<2\log\Lambda_0$. For $a_j=\wcoef{T_j(\varphi)}$, we have $a_0=1$, $a_1=\wcoef{\varphi}\le\Lambda_\varphi$, and inductively $a_{j+1}\le 2\wcoef{\varphi}a_j+a_{j-1}\le (2\wcoef{\varphi}+1)\Lambda_\varphi^j = \Lambda_\varphi^{j+1}$ by submultiplicativity, so $a_h\le\Lambda_\varphi^h$. Since $\wcoef{\prod_{j=1}^r(1-x/j)} \le \prod_{j=1}^r(1+N/j)\le\Lambda_0^r$ and $T_h(\varphi(0))\ge 1$, submultiplicativity yields
\[
  \log \wcoef{q}
  \le r\log\Lambda_0 + h\log\Lambda_\varphi
  \le \sqrt{Rr}\log\Lambda_0 + 10\sqrt{Rr}\log^2\Lambda_0
  \le 11\sqrt{Rr}\log^2\Lambda_0.
\]

Finally, let $K$ be the product projection underlying $D$, so $D=\sum_{i\in\supp K}Q_i$ with single-qubit rank-$1$ projectors $Q_i=I-K_i$. Expanding $D^j$ yields at most $N^j$ terms $Q_{i_1}\cdots Q_{i_j}$ (with $D^0 = I$), each simplifying by commutativity and idempotence to the product projection onto mismatches on the at most $j$ distinct qubits in $(i_1,\dots,i_j)$; weighting by $q_j$ and summing over $j$ gives support at most $\deg q$ and total $\ell_1$ weight at most $\sum_j |q_j|N^j = \wcoef{q}$.
\end{proof}

\subsection{Auxiliary Spectral Separation via Pauli Twirling on Register \texorpdfstring{$X$}{X}}\label{sec:twirl}

For an $N$-qubit $\QAC$ circuit $V$ on $X \otimes A$ and a mismatch counter $D$, define the pullback projectors
\begin{equation}
  \PhiZero = V^\ad [D = 0] V, \qquad \PhiHigh = V^\ad [D \ge R] V.
  \label{eq:pullback}
\end{equation}
We first bound how strongly any Pauli string on $X$ can couple $\PhiZero$ to $\PhiHigh$.

\begin{lemma}[Pauli Transition Bound]\label{lem:pauli-transition}
Fix $d \in \mathbb{Z}_{\ge 0}$, $\kw \ge 1$, and $0 \le \zeta < \bhi$, and set $R = \lceil m^{\bhi} \rceil$. For all sufficiently large $m \ge m_2(d, \kw, \zeta, \bhi)$, every $N$-qubit $\QAC$ circuit $V$ of depth at most $d$ with $m \le N \le m^{\kw}$, every mismatch counter $D$, and every $\sigma \in \Pauli$,
\begin{equation}
  \norm{\PhiHigh \hsigma \PhiZero} \le \kappa,
  \qquad \kappa = 2^m e^{-m^\zeta}.
  \label{eq:transition}
\end{equation}
\end{lemma}

\begin{proof}
Let $J \subseteq [m]$ be the set of non-identity tensor factors of $\sigma$, and write each $\sigma_i = P_{i,+} - P_{i,-}$ ($i \in J$) in terms of its rank-$1$ eigenprojectors. For $a \in \{-1,1\}^J$, let $K_a$ act as $P_{i,a_i}$ on $i \in J$ and as $I$ elsewhere (including on $A$), so $\hsigma = \sum_{a \in \{-1,1\}^J} (\prod_{i\in J} a_i) K_a$ has $2^{|J|} \le 2^m$ terms.

Fix $C = \lfloor\max(\kw,\bhi)\rfloor + 1$, pad with $N_* - N$ idle qubits to width $N_* = m^C \ge N$, and write $V_* = V \otimes I$, $D_* = D \otimes I$, and $[M_a = 0] = K_a \otimes I$ for the mismatch counter $M_a$ of $K_a \otimes I$. Setting $\theta_1 = \zeta/C < \bhi/C = \theta_2$ gives $N_*^{\theta_1} = m^\zeta$ and $N_*^{\theta_2} = m^{\bhi}$, and integrality of $\operatorname{spec}(D)$ gives $[D_* \ge N_*^{\theta_2}] = [D \ge R] \otimes I$. For $m \ge \max\{1, \lceil N_0(d,\theta_1,\theta_2)^{1/C}\rceil\}$, unitary invariance and Theorem~\ref{th:localization} at width $N_*$ yield
\[
  \norm{\PhiHigh K_a \PhiZero}
  = \norm{[D\ge R]\,V K_a V^\ad[D=0]}
  = \norm{[D_*\ge N_*^{\theta_2}]\,V_*[M_a=0]V_*^\ad[D_*=0]}
  \le e^{-m^\zeta}.
\]
Summing over the at most $2^m$ terms via the triangle inequality gives~\eqref{eq:transition}.
\end{proof}

Averaging $\PhiZero$ over $\Pauli$ removes all dependence on $X$ while preserving~\eqref{eq:transition}, allowing a spectral cutoff on $A$ to separate $\PhiZero$ from $\PhiHigh$.

\begin{lemma}[Auxiliary Separating Projector]\label{lem:selector}
Fix $d \in \mathbb{Z}_{\ge 0}$, $\kw \ge 1$, and $1 < \sloc < \zeta < \bhi$, and set $R = \lceil m^{\bhi} \rceil$. For all sufficiently large $m \ge m_3(d, \kw, \sloc, \zeta, \bhi)$, every $N$-qubit $\QAC$ circuit $V$ of depth at most $d$ with $m \le N \le m^{\kw}$, and every mismatch counter $D$, there exists an orthogonal projector $\pi$ on $A$ such that $\Pi = I_X \otimes \pi$ satisfies
\begin{equation}
  \norm{(I - \Pi)\PhiZero} \le e^{-m^{\sloc}},
  \qquad \norm{\PhiHigh \Pi} \le e^{-m^{\sloc}}.
  \label{eq:selector}
\end{equation}
\end{lemma}

\begin{proof}
Define the Pauli-twirled operator
\[
  \mathcal{T} = 4^{-m}\sum_{\sigma \in \Pauli} \hsigma \PhiZero \hsigma^\ad,
\]
and expand $\PhiZero=\sum_{\sigma'\in\Pauli}\sigma'\otimes(\PhiZero)_{\sigma'}$. Conjugation by $\hsigma$ fixes the $\sigma'=I_X$ term. For any $\sigma' \in \Pauli \setminus \{I_X\}$, pick a weight-$1$ Pauli $v \in \Pauli$ anticommuting with $\sigma'$, so $v\sigma'v^\ad = -\sigma'$; right-multiplying $\sigma$ by $v$ permutes $\Pauli$ up to phase and cancels the $\sigma$ and $\sigma v$ contributions to $\sigma'\otimes(\PhiZero)_{\sigma'}$. Hence
\begin{equation}
  \mathcal{T} = I_X \otimes S_A,
  \qquad S_A = 2^{-m}\tr_X(\PhiZero),
  \label{eq:twirl}
\end{equation}
where $0 \preceq S_A \preceq I$ because $\mathcal{T}$ is a convex combination of orthogonal projectors.

Retaining only the $\sigma = I_X$ term in $\mathcal{T}$ gives
\begin{equation}
  \PhiZero \preceq 4^m \mathcal{T}.
  \label{eq:loewner-low}
\end{equation}
In the opposite direction, using $\PhiZero^2 = \PhiZero$ and $\PhiHigh^2 = \PhiHigh$, we write
\[
  \PhiHigh \mathcal{T} \PhiHigh
  = 4^{-m}\sum_{\sigma \in \Pauli} (\PhiHigh \hsigma \PhiZero)(\PhiHigh \hsigma \PhiZero)^\ad.
\]
For every vector $\ket{\psi}$ and every $\sigma \in \Pauli$, Lemma~\ref{lem:pauli-transition} implies
\[
  \bra{\psi}(\PhiHigh \hsigma \PhiZero)(\PhiHigh \hsigma \PhiZero)^\ad \ket{\psi}
  = \norm{\PhiZero \hsigma^\ad \PhiHigh (\PhiHigh \ket{\psi})}^2
  \le \norm{\PhiHigh \hsigma \PhiZero}^2 \norm{\PhiHigh \ket{\psi}}^2
  \le \kappa^2 \bra{\psi}\PhiHigh\ket{\psi},
\]
so averaging over $\sigma \in \Pauli$ yields
\begin{equation}
  \PhiHigh \mathcal{T} \PhiHigh \preceq \kappa^2 \PhiHigh.
  \label{eq:high}
\end{equation}

Set $\tcut = e^{-m^\zeta}$, $\pi = [S_A \ge \tcut]$, $\Pi = I_X \otimes \pi$, and $\PiPerp = I - \Pi$. Spectral decomposition of $S_A$ and~\eqref{eq:twirl} give $\PiPerp\mathcal{T}\PiPerp \preceq \tcut\PiPerp$ and $\tcut\Pi \preceq \mathcal{T}$. Combining these bounds with~\eqref{eq:loewner-low} and~\eqref{eq:high} via $\norm{Y}^2 = \norm{YY^\ad}$ yields
\begin{align*}
  \norm{\PiPerp \PhiZero}^2 &= \norm{\PiPerp \PhiZero \PiPerp} \le 4^m \norm{\PiPerp \mathcal{T} \PiPerp} \le 4^m \tcut = 4^m e^{-m^\zeta}, \\
  \norm{\PhiHigh \Pi}^2 &= \norm{\PhiHigh \Pi \PhiHigh} \le \tcut^{-1}\norm{\PhiHigh \mathcal{T} \PhiHigh} \le \kappa^2 / \tcut = 4^m e^{-m^\zeta}.
\end{align*}
Taking square roots bounds both $\norm{\PiPerp \PhiZero}$ and $\norm{\PhiHigh \Pi}$ by $2^m e^{-m^\zeta/2}$. Since $\zeta > \sloc > 1$, we have $\frac{1}{2}m^\zeta \ge m\log 2 + m^{\sloc}$ for all sufficiently large $m$, which gives~\eqref{eq:selector}.
\end{proof}

\subsection{Low-Support Polynomial Replacement}\label{sec:replacement}

Combining the auxiliary projector $\Pi$ of Lemma~\ref{lem:selector} with the Chebyshev filter of Lemma~\ref{lem:polynomial} replaces $V^\ad[D=0]V$ by a low-support linear combination of conjugated product projections.

\begin{lemma}[Low-Support Polynomial Replacement]\label{lem:replacement}
Fix $d \in \mathbb{Z}_{\ge 0}$, $\kw \ge 1$, and $\bhi > 1$. Set $R = \lceil m^{\bhi} \rceil$, $\sloc = 1 + (\bhi - 1)/3$, $\zeta = 1 + 2(\bhi - 1)/3$, and let $m \ge m_3(d, \kw, \sloc, \zeta, \bhi)$. For any integers $m \le N \le m^{\kw}$ and $2 \le r \le R/2$ with $b(r) := 11\sqrt{Rr}\log^2(2 + N + R) \le m^{\sloc}$ and $3r + 2 \le m^{\sloc}$, every $N$-qubit $\QAC$ circuit $V$ of depth at most $d$, and every mismatch counter $D$, there exist an orthogonal projector $\pi$ on $A$ ($\Pi = I_X \otimes \pi$), product projections $K_\iota$ of support at most $b(r)$, and real coefficients $t_\iota$ such that
\begin{equation}
  \norm{V^\ad [D = 0] V - \sum_\iota t_\iota \Pi V^\ad K_\iota V \Pi} \le e^{-2r},
  \qquad \sum_\iota |t_\iota| \le e^{b(r)}.
  \label{eq:replacement}
\end{equation}
\end{lemma}

\begin{proof}
Let $q$ be the real polynomial from Lemma~\ref{lem:polynomial} for $(N, R, r)$. Since $\Lambda_0 = 2 + N + R \ge 5$ implies $\log \Lambda_0 > 1$,~\eqref{eq:filter-weight} gives both $\deg q \le b(r)$ and $\wcoef{q} \le e^{b(r)}$. Let $\Pi = I_X \otimes \pi$ be the projector from Lemma~\ref{lem:selector} for $\PhiZero = V^\ad [D=0] V$ and $\PhiHigh = V^\ad [D \ge R] V$, and set $\Qflt = V^\ad q(D) V$.

Because $\PhiZero, \Qflt, \PhiHigh$ are all functions of $V^\ad D V$, the projector $\PhiHigh$ commutes with $\PhiZero - \Qflt$, so the cross terms between $I - \PhiHigh$ and $\PhiHigh$ vanish:
\[
  \PhiZero - \Pi \Qflt \Pi
  = (\PhiZero - \Pi \PhiZero \Pi)
  + \Pi (I - \PhiHigh)(\PhiZero - \Qflt)(I - \PhiHigh)\Pi
  + \Pi \PhiHigh (\PhiZero - \Qflt) \PhiHigh \Pi.
\]
We bound each term in operator norm:
\begin{enumerate}
  \item[(i)] $\PhiZero-\Pi\PhiZero\Pi=\PiPerp\PhiZero+\Pi\PhiZero\PiPerp$ has norm at most $2\norm{\PiPerp\PhiZero}\le 2e^{-m^{\sloc}}$ by~\eqref{eq:selector};
  \item[(ii)] on $[V^\ad D V < R]$, $\PhiZero - \Qflt$ has eigenvalues $\delta_{j,0} - q(j)$ ($0 \le j < R$), which vanish at $j=0$ and have magnitude at most $e^{-3r}$ for $1 \le j < R$ by~\eqref{eq:filter-error}, so the middle term has norm at most $e^{-3r}$;
  \item[(iii)] since $R\ge 1$, $\PhiHigh\PhiZero=0$, so the third term equals $-(\PhiHigh\Pi)^\ad\Qflt(\PhiHigh\Pi)$ and has norm at most $\norm{\PhiHigh\Pi}^2\norm{\Qflt}\le e^{b(r)-2m^{\sloc}}$ by~\eqref{eq:selector} and $\norm{\Qflt}\le\wcoef{q}\le e^{b(r)}$.
\end{enumerate}
Combining these bounds with $b(r) \le m^{\sloc}$, $m^{\sloc} \ge 3r + 2$, and $r \ge 2$ yields
\begin{equation}
  \norm{\PhiZero - \Pi \Qflt \Pi}
  \le 2e^{-m^{\sloc}} + e^{-3r} + e^{b(r) - 2m^{\sloc}}
  \le 3e^{-m^{\sloc}} + e^{-3r}
  \le (3e^{-2} + e^{-2})e^{-2r}
  < e^{-2r}.
  \label{eq:three-term}
\end{equation}
Expanding $q(D) = \sum_\iota t_\iota K_\iota$ via Lemma~\ref{lem:polynomial} into product projections of support at most $\deg q \le b(r)$ with $\sum_\iota |t_\iota| \le \wcoef{q} \le e^{b(r)}$ completes the proof.
\end{proof}

\subsection{Backward Layer-by-Layer Peeling: Proof of Theorem~\ref{th:approximation}}\label{sec:induction}

\begin{proof}[Proof of Theorem~\ref{th:approximation}]
Pad $U$ with empty Toffoli layers to depth $d$, and factor $U$ from output to input as $V_0 = U$ and $V_{j-1} = G_j V_j$ ($1 \le j \le d$), where $G_j$ comprises the $j$-th generalized Toffoli layer from the output together with its adjacent output-side single-qubit layer. Each $V_j$ then has depth $d - j \le d$, and $V_d$ consists solely of single-qubit unitaries.

\paragraph{Peeling invariant and unified budget.}
We construct Hermitian approximants $E_0, E_1, \dots, E_d$ of the form
\begin{equation}
  E_j = \sum_\iota c_\iota (I_X \otimes C_\iota)\,
        V_j^\ad K_\iota V_j\, (I_X \otimes C_\iota^\ad),
  \qquad c_\iota \in \mathbb{R},\quad \norm{C_\iota} \le 1,
  \label{eq:invariant}
\end{equation}
where each $K_\iota$ is a product projection on $N$ qubits, each $C_\iota$ is a contraction on $A$, and the budget $s_j$ bounds
\[
  \text{(i)}\quad |\supp K_\iota| \le s_j \;\;\text{for all } \iota,
  \qquad\text{and}\qquad
  \text{(ii)}\quad \sum_\iota |c_\iota| \le e^{s_j}.
\]
Fix $\tau = \frac{1}{2}\min\{\eta, (1-\alpha)2^{-d}\} > 0$, and set $\bhi = 1 + \tau$, $R = \lceil m^{\bhi} \rceil$, $\sloc = 1 + \tau/3$, $\zeta = 1 + 2\tau/3$, and $\Lambda_0 = 2 + N + R$. Define $s_0, \dots, s_d$ starting from $s_0 = m^\alpha + \log d$ by
\begin{equation}
  r_j = \lceil 3 s_{j-1} \rceil,
  \qquad
  s_j = 4 s_{j-1} + 11\sqrt{R r_j}\,\log^2 \Lambda_0
  \qquad (1 \le j \le d).
  \label{eq:budget-rec}
\end{equation}

To bound $s_j$ independently of $N$, set $\overline{\Lambda}_0=2+m^{\kw}+2m^{1+\tau}\ge\Lambda_0$ and $\overline{M}=4+22\log^2\overline{\Lambda}_0$, and define $u_0=s_0$ and $u_j=\overline{M}\sqrt{R u_{j-1}}$ ($1\le j\le d$). Unrolling this recurrence by induction gives
\begin{equation}
  u_j = \overline{M}^{2(1 - 2^{-j})}\, R^{1 - 2^{-j}} s_0^{2^{-j}}
  \qquad (0 \le j \le d).
  \label{eq:budget-closed}
\end{equation}
For $m^\alpha \ge \log d$, we have $s_0 \le 2m^\alpha$ and $R \le 2m^{1+\tau}$, so~\eqref{eq:budget-closed} gives $u_j \le 2\overline{M}^2 m^{\theta_j}$ with $\theta_j = (1+\tau)(1-2^{-j}) + \alpha 2^{-j}$. Since $\theta_j$ increases with $j$ to
\[
  \theta_d = 1-(1-\alpha)2^{-d}+\tau(1-2^{-d})
  \le 1-\tfrac{1}{2}(1-\alpha)2^{-d} < 1 < \sloc < \bhi,
\]
and since $\tau(1-2^{-d}) < \eta/2$ and $\overline{M} = O_{\kw,\tau}(\log^2 m)$, there exists $m_0 = m_0(d,\kw,\alpha,\eta)$ such that for all $m \ge m_0$:
\begin{enumerate}
  \item[(i)] $m \ge \max\{5, (\log d)^{1/\alpha}, m_3(d,\kw,\sloc,\zeta,\bhi)\}$;
  \item[(ii)] $u_j \le 2\overline{M}^2 m^{\theta_d} \le \min\{\frac{1}{20}m^{\sloc}, \frac{1}{20}R\}$ for all $0 \le j \le d$;
  \item[(iii)] $u_d \le m^{1 - (1-\alpha)2^{-d} + \eta}$.
\end{enumerate}

Clearly $s_0 = u_0 \ge 1$. If $1 \le s_{j-1} \le u_{j-1}$, then $s_{j-1} \le R$ and $r_j \le 4s_{j-1}$, so~\eqref{eq:budget-rec} and $s_{j-1} \le \sqrt{R s_{j-1}}$ give
\[
  1 \le s_j
  \le (4+22\log^2\overline{\Lambda}_0)\sqrt{R s_{j-1}}
  \le \overline{M}\sqrt{R u_{j-1}} = u_j.
\]
Hence $1 \le s_j \le u_j$ for all $0 \le j \le d$. By (ii), each $1 \le j \le d$ satisfies $3 \le r_j \le R/5 \le R/2$, $b(r_j) \le s_j \le m^{\sloc}/20 \le m^{\sloc}$, and $3r_j + 2 \le \frac{3}{5}m^{\sloc} + 2 \le m^{\sloc}$, verifying all hypotheses of Lemma~\ref{lem:replacement}.

\paragraph{Inductive step.}
At $j=0$, $E_0=E$ satisfies~\eqref{eq:invariant} with a single term ($c=1$, $K=\Pout$, $C=I_A$), as $|\supp \Pout|=1\le s_0$. Suppose~\eqref{eq:invariant} holds at $j-1$. For each $K_\iota$, Lemma~\ref{lem:layer} expands $G_j^\ad K_\iota G_j = \sum_\ell a_{\iota\ell} P_{\iota\ell}$ with $a_{\iota\ell} \in \mathbb{R}$ and $\sum_\ell |a_{\iota\ell}| \le 18^{s_{j-1}}$. Using $V_{j-1} = G_j V_j$, we have
\[
  E_{j-1}
  = \sum_{\iota,\ell} c_\iota a_{\iota\ell}
        (I_X\otimes C_\iota)V_j^\ad P_{\iota\ell}V_j(I_X\otimes C_\iota^\ad),
  \qquad
  \sum_{\iota,\ell}|c_\iota a_{\iota\ell}| \le e^{s_{j-1}} 18^{s_{j-1}} \le e^{4s_{j-1}},
\]
since $1 + \log 18 < 4$. Applying Lemma~\ref{lem:replacement} (at depth $d - j \le d$) with parameter $r_j$ to the mismatch counter of each $P_{\iota\ell}$ approximates $V_j^\ad P_{\iota\ell} V_j$ within $e^{-2r_j}$ in operator norm by
\[
  \sum_h t_{\iota\ell h}\,(I_X\otimes\pi_{\iota\ell})
       V_j^\ad K_{\iota\ell h}V_j (I_X\otimes\pi_{\iota\ell}),
  \qquad
  \sum_h|t_{\iota\ell h}|\le e^{b(r_j)}.
\]
Substituting these sums into $E_{j-1}$ defines $E_j$: the term indexed by $(\iota,\ell,h)$ has real weight $c_\iota a_{\iota\ell} t_{\iota\ell h}$, product projection $K_{\iota\ell h}$ of support at most $b(r_j) \le s_j$, and outer contraction $C_\iota \pi_{\iota\ell}$ with $\norm{C_\iota \pi_{\iota\ell}} \le 1$. Its total $\ell_1$ weight is at most
\[
  \sum_{\iota,\ell}|c_\iota a_{\iota\ell}|\sum_h |t_{\iota\ell h}|
  \le e^{4s_{j-1} + b(r_j)} = e^{s_j},
\]
establishing (i) and (ii) for $E_j$. Since $\norm{I_X \otimes C_\iota} \le 1$, $r_j \ge 3s_{j-1}$, and $s_{j-1} \ge s_0$, the single-step error satisfies
\[
  \norm{E_j-E_{j-1}}
  \le \sum_{\iota,\ell}|c_\iota a_{\iota\ell}|e^{-2r_j}
  \le e^{4s_{j-1}-2r_j}
  \le e^{-2s_{j-1}}\le e^{-2s_0}.
\]

\paragraph{Termination and Pauli degree on $X$.}
Summing over $j = 1, \dots, d$ and recalling $s_0 = m^\alpha + \log d$ gives
\[
  \norm{E - E_d}
  \le \sum_{j=1}^{d} \norm{E_j - E_{j-1}}
  \le d e^{-2s_0}
  = d^{-1} e^{-2m^\alpha}
  \le e^{-m^\alpha}.
\]
Because $V_d$ is a tensor product of single-qubit unitaries, each $V_d^\ad K_\iota V_d$ is a product projection with the same support size $|\supp K_\iota| \le s_d$; its Pauli expansion on $X$ acts as $I$ outside $\supp K_\iota$ and thus has degree at most $s_d$, which conjugation of the $A$-factors by $C_\iota$ preserves. Setting $B = E_d$ yields a Hermitian operator with $\norm{E - B} \le e^{-m^\alpha}$ and $\deg_X(B) \le s_d \le u_d \le m^{1-(1-\alpha)2^{-d}+\eta}$.
\end{proof}

\section{From Operator Approximation to \texorpdfstring{$\AC \not\subseteq \QAdv$}{AC0 not in BQAC0/qpoly}}\label{sec:separation}

\subsection{Approximate Degree Upper Bound: Proof of Corollary~\ref{cor:degree}}

For any classical input $x \in \{0,1\}^m$ on $X$ and $x$-independent density operator $\rho$ on $A$, the acceptance probability is $p_{\mathrm{acc}}(x) = \tr((\ket{x}\bra{x}\otimes\rho)E)$. Given a Hermitian operator $B$ on $X \otimes A$, let $Z_J \in \Pauli$ act as $\mathrm{Z}$ on $J \subseteq [m]$ and as $I$ elsewhere. Since $\mathrm{X}$ and $\mathrm{Y}$ have zero computational-basis diagonal and $\bra{x_i}\mathrm{Z}\ket{x_i} = 1 - 2x_i$,
\begin{equation}
  p_\rho(x):=\tr\bigl((\ket{x}\bra{x}\otimes\rho)B\bigr)
   =\sum_{J\subseteq[m],\,|J|\le\deg_X(B)}\tr(\rho B_{Z_J})\prod_{i\in J}(1-2x_i).
  \label{eq:classical-evaluation}
\end{equation}
Since $B = B^\ad$ and $Z_J = Z_J^\ad$, uniqueness of the Pauli expansion gives $B_{Z_J}^\ad = B_{Z_J}$ and $\tr(\rho B_{Z_J}) \in \mathbb{R}$, so $p_\rho$ is a real polynomial of degree at most $\deg_X(B)$ satisfying $|p_\rho(x) - p_{\mathrm{acc}}(x)| \le \norm{E - B}$ for all $x \in \{0,1\}^m$.

\begin{proof}[Proof of Corollary~\ref{cor:degree} from Theorem~\ref{th:approximation}]
Set $\alpha = \min\{1/2, 2^{d-1}\gamma\}$ and $\eta = \gamma/3$, so the exponent in~\eqref{eq:degree-bound} is at most $1 - 2^{-d} + \frac{5}{6}\gamma < 1 - 2^{-d} + \gamma$. For $m \ge m_1(d,\kw,\gamma) := \max\{m_0(d,\kw,\alpha,\eta), \lceil(\log 100)^{1/\alpha}\rceil\}$, Theorem~\ref{th:approximation} and~\eqref{eq:classical-evaluation} give a real polynomial $p_\rho$ of degree at most $m^{1-2^{-d}+\gamma}$ such that, with $\varepsilon = e^{-m^\alpha} \le 1/100$,
\[
  |p_\rho(x)-f(x)|
  \le |p_\rho(x)-p_{\mathrm{acc}}(x)|+|p_{\mathrm{acc}}(x)-f(x)|
  \le \varepsilon + 1/3,
\]
proving the first inequality in~\eqref{eq:cor-degree}.

Since $p_{\mathrm{acc}}(x) \in [0,1]$, we have $p_\rho(x) \in [-\varepsilon, 1+\varepsilon]$, so $\tilde{p}(x) = (p_\rho(x)+\varepsilon)/(1+2\varepsilon)$ lies in $[0,1]$ and satisfies
\[
  |\tilde{p}(x)-f(x)| \le \frac{1/3+2\varepsilon}{1+2\varepsilon}
  \le \frac{53}{153} < \frac{3}{8}
\]
on both $0$- and $1$-inputs. Because $\mathrm{maj}_3(\xi) = 3\xi^2 - 2\xi^3$ has derivative $6\xi(1-\xi) \ge 0$ on $[0,1]$, satisfies $\mathrm{maj}_3(3/8) = 81/256 < 1/3$, and obeys $\mathrm{maj}_3(1-\xi) = 1 - \mathrm{maj}_3(\xi)$, the composition $\mathrm{maj}_3 \circ \tilde{p}$ has degree at most $3\deg p_\rho$ and approximates $f$ within $1/3$.
\end{proof}

\subsection{Monotone Universal Selector CNF and Proof of Theorem~\ref{th:separation}}\label{sec:universal-cnf}

We use Sherstov's near-linear approximate degree lower bound for monotone constant-width formulas.

\begin{theorem}[Sherstov~{\cite[Theorem~5.1]{She25}}]\label{th:sherstov}
For every fixed $\delta \in (0,1)$, there exist a constant $c_\delta > 0$, a positive integer width $w = w(\delta)$, and a family of $m$-variate Boolean functions $F_m \colon \{0,1\}^m \to \{0,1\}$ each computable by a monotone CNF formula of width at most $w$, such that for all sufficiently large $m$,
\begin{equation}
  \ade_{1/2 - m^{-2}}(F_m) \ge c_\delta m^{1 - \delta}.
  \label{eq:sherstov}
\end{equation}
\end{theorem}

Sherstov~\cite[Theorem~5.1]{She25} proves~\eqref{eq:sherstov} for monotone width-$w$ DNFs $g_m(x)=\bigvee_{C\in\mathcal{C}}\bigwedge_{i\in C}x_i$; dualizing to $F_m(x)=1-g_m(1-x)=\bigwedge_{C\in\mathcal{C}}\bigvee_{i\in C}x_i$ yields monotone width-$w$ CNFs with identical approximate degree at every error, since the involution $p(x) \mapsto 1 - p(1 - x)$ preserves polynomial degree and uniform error on $\{0,1\}^m$.

Every $m$-variate width-$w$ CNF has approximate degree $O_w(m^{1 - 1/(w+1)})$~\cite[Theorem~1.2]{She20}. To apply the degree comparison at all constant depths, we construct a single $\PClass$-uniform monotone depth-$2$ CNF family $H = (H_n)_{n \ge 1}$ that, for every fixed $w$, embeds every monotone width-$w$ CNF on $m_w(n) = \Theta_w(n^{1/w})$ variables at every sufficiently large length $n$.

\paragraph{Monotone universal selector CNF for width-$w$ formulas.}
For integers $m, w \ge 1$, let $\cL_{m,w} = (\{0, x_1, \dots, x_m\})^w$ be the set of all $(m+1)^w$ tuples of length $w$ over the constant $0$ and the $m$ positive literals on $x = (x_1, \dots, x_m)$. Introducing a selector variable $y_{\vec{\ell}}$ for each $\vec{\ell} = (\ell_1, \dots, \ell_w) \in \cL_{m,w}$, define the \emph{monotone universal selector CNF for width-$w$ formulas} by
\begin{equation}
  H_{m,w}(x, y) = \bigwedge_{\vec{\ell} = (\ell_1, \dots, \ell_w) \in \cL_{m,w}}
    \bigl(y_{\vec{\ell}} \lor \ell_1(x) \lor \cdots \lor \ell_w(x)\bigr),
  \label{eq:universal}
\end{equation}
omitting zero entries from the OR gate. Each clause of $H_{m,w}$ is monotone and has width at most $w+1$ (including $y_{\vec{\ell}}$). Given any width-$w$ monotone CNF $F$ on $x$, list the distinct variables of each clause in increasing order and pad with zeros to form a subset $\mathcal{T} \subseteq \cL_{m,w}$; setting $y_{\vec{\ell}} = 0$ for $\vec{\ell} \in \mathcal{T}$ and $y_{\vec{\ell}} = 1$ otherwise activates the clauses in $\mathcal{T}$ and satisfies all others, realizing $H_{m,w}(x,y) = F(x)$.

\paragraph{Disjoint-block construction for all lengths $n$.}
For input length $n \ge 1$, allocate consecutive disjoint blocks in $[n]$ of sizes $B_w = \lfloor n/2^w \rfloor$ ($w \ge 1$), which fit since $\sum_{w=1}^\infty B_w \le n$. Call block $w$ \emph{active} if $B_w \ge 1 + 2^w$, let $m_w = m_w(n) \ge 1$ be the largest integer satisfying
\begin{equation}
  m_w + (m_w + 1)^w \le B_w
  \label{eq:block-size}
\end{equation}
(setting $m_w = 0$ when $B_w < 1 + 2^w$), and assign the first $m_w$ bits of block $w$ to $x^{(w)}$ and the next $M_w = (m_w+1)^w$ bits to $y^{(w)}$ for $H_{m_w,w}$. Letting $\mathcal{U}_n \subseteq [n]$ denote all remaining unused input indices, define
\[
  H_n(z)
  = \bigwedge_{w\text{ active}} H_{m_w,w}\bigl(x^{(w)},y^{(w)}\bigr)
    \ \land\! \bigwedge_{i\in\mathcal{U}_n} z_i
\]
(for $n < 6$, no block is active since $2(1+2) = 6$, so $H_n(z) = \bigwedge_{i=1}^n z_i$).

\paragraph{Linear size, variable relevance, $\PClass$-uniformity, and block growth.}
Each active block $w$ contributes at most $(w + 1) M_w$ literal occurrences, while the unit clauses contribute $|\mathcal{U}_n| = n - \sum_{w\text{ active}}(m_w + M_w)$ literals. Since $M_w < m_w + M_w \le B_w$, the total literal count of $H_n$ satisfies
\[
  n - \sum_{w\text{ active}}(m_w + M_w) + \sum_{w\text{ active}}(w + 1) M_w
  \le n + \sum_{w=1}^{\infty} w B_w
  \le n + n \sum_{w=1}^{\infty} \frac{w}{2^w}
  = 3n.
\]

Every variable of $H_n$ can be isolated by a restriction: for $i \in \mathcal{U}_n$, set all other variables to $1$; for $x_i^{(w)}$, set all unit-clause bits and all selectors except that of $(x_i^{(w)},0,\dots,0)$ to $1$ and the remaining selector to $0$; and for $y^{(w)}_{\vec{\ell}}$, set $x^{(w)} = 0$ and all unit-clause bits and other selectors to $1$.

On input $1^n$, one computes $B_w$ for $1 \le w \le \lfloor\log_2 n\rfloor$, finds $m_w$ by binary search using $O(\log n)$-bit arithmetic (capping intermediate powers at $B_w + 1$), enumerates $\cL_{m_w,w}$ lexicographically to emit the clauses of each active block, and outputs the unit clauses on $\mathcal{U}_n$. Since tuples have length at most $\log_2 n$ and $\sum_{w\text{ active}} M_w \le n$, this runs in time $\poly(n)$.

Finally, fix any constant $w \ge 1$. For all $n \ge 2^w(1 + 2^w)$, we have $B_w \ge 1 + 2^w$ (so $m_w \ge 1$) and $n/2^{w+1} \le B_w \le n/2^w$. By maximality of $m_w$ in~\eqref{eq:block-size} and $m_w + 1 \le 2m_w \le (3m_w)^w$, $m_w + 2 \le 3m_w$,
\[
  m_w^w < B_w < (m_w + 1) + (m_w + 2)^w \le 2(3m_w)^w.
\]
Substituting $n/2^{w+1} \le B_w \le n/2^w$ and taking $w$-th roots gives
\begin{equation}
  \frac{1}{3}\!\left(\frac{n}{2^{w+2}}\right)^{1/w} < m_w(n) < \frac{n^{1/w}}{2},
  \qquad\text{so}\qquad
  m_w(n) = \Theta_w(n^{1/w})
  \label{eq:mw-growth}
\end{equation}
for all $n \ge 2^w(1 + 2^w)$; in particular, $m_w(n) \to \infty$ and $n < 4 \cdot 6^w m_w(n)^w$ as $n \to \infty$.

\begin{proof}[Proof of Theorem~\ref{th:separation} from Theorem~\ref{th:approximation}]
Fix any depth $d \ge 1$ and width exponent $c_0 \ge 1$. Set $\delta = 2^{-d}/3 \in (0, 1/6]$, and let $w = w(\delta)$, $c_\delta > 0$, and the monotone width-$w$ CNFs $(F_m)$ be given by Theorem~\ref{th:sherstov}. Set $\gamma_d = 2/w > 0$, $\alpha = 1/6$, $\eta = \delta/2 = 2^{-d}/6$, and $\kw = w c_0 + 1$, so the degree exponent in Theorem~\ref{th:approximation} equals
\[
  1 - (1 - \alpha)2^{-d} + \eta
  = 1 - \tfrac{5}{6}\cdot 2^{-d} + \tfrac{1}{6}\cdot 2^{-d}
  = 1 - 2\delta.
\]
Write $m = m_w(n)$. By~\eqref{eq:mw-growth}, $m \to \infty$ and $n < 4\cdot 6^w m^w$, so for all sufficiently large $n$, block $w$ is active, $(n+1)^{c_0} \le (4\cdot 6^w+1)^{c_0} m^{wc_0} \le m^{\kw}$, $e^{-m^{1/6}} \le m^{-2}$, and Theorems~\ref{th:approximation} and~\ref{th:sherstov} apply at length $m$. Suppose a depth-$d$ circuit $U_n$ of width $N_n \le (n+1)^{c_0}$ with advice $\varrho_n$ succeeds on $H_n(z)$ with probability at least $1/2 + 2m^{-2}$ for all $z \in \{0,1\}^n$.

Let $z(x) \in \{0,1\}^n$ be the restriction that programs $y^{(w)}$ to realize $F_m$ on $x = x^{(w)} \in \{0,1\}^m$, sets $y^{(w')} = 1$ and $x^{(w')} = 0$ in all other active blocks $w' \ne w$, and sets $z_i = 1$ for $i \in \mathcal{U}_n$, so that $H_n(z(x)) = F_m(x)$. Taking the $m$ free-input qubits of $x^{(w)}$ as $X$ and combining the fixed input bits, clean ancillas, and $\varrho_n$ into a state $\rho_n$ on $A$, we have $m \le n \le N_n \le m^{\kw}$ and $|p_{\mathrm{acc}}(z(x)) - F_m(x)| \le 1/2 - 2m^{-2}$. Applying Theorem~\ref{th:approximation} and~\eqref{eq:classical-evaluation} yields a real polynomial $p_{\rho_n}$ of degree at most $m^{1-2\delta}$ satisfying
\[
  \max_{x \in \{0,1\}^m} |p_{\rho_n}(x) - F_m(x)|
  \le \tfrac{1}{2} - 2m^{-2} + e^{-m^{1/6}}
  \le \tfrac{1}{2} - m^{-2}.
\]
Thus $c_\delta m^{1-\delta} \le \ade_{1/2-m^{-2}}(F_m) \le m^{1-2\delta}$, contradicting $m \to \infty$. Every such circuit therefore has worst-case success probability at most $1/2 + 2m_w(n)^{-2} < 1/2 + 18\cdot 2^{2(w+2)/w} n^{-2/w} = 1/2 + O_d(n^{-\gamma_d})$ by~\eqref{eq:mw-growth}. Since $H \in \AC$, it follows that $\AC \not\subseteq \QAdv$.
\end{proof}

\section{Applications to Quantum States}\label{sec:quantum-applications}

We next apply Theorem~\ref{th:approximation} to distinguishing state $2$-designs from Haar-random states and to preparing low-energy states of NLTS Hamiltonians, using a common flag-qubit reduction to convert multi-qubit computational-basis pattern tests into single-qubit acceptance.

\paragraph{Testing an output pattern with a flag qubit.}
Let $U$ be an $N$-qubit, depth-$d$ $\QAC$ circuit, let $S$ be an ordered subset of $s \ge 1$ output qubits, and fix $y \in \{0,1\}^s$, with pattern projector $P_y = \ket{y}\bra{y}_S \otimes I_{\bar S}$. Append a flag qubit $F$ initialized in $\ket{0}$, let $T_{S,F} = (I-P_{1^s})\otimes I_F + P_{1^s}\otimes \mathrm{X}_F$ be the generalized Toffoli with controls $S$ and target $F$, and let $R_y = R_y^\ad$ apply $\mathrm{X}$ to each qubit of $S$ where $y$ has a zero bit, so $R_y P_{1^s} R_y = P_y$. Defining
\[
  T_y := (R_y\otimes I_F) T_{S,F} (R_y\otimes I_F) = (I-P_y)\otimes I_F + P_y\otimes \mathrm{X}_F,
  \qquad U_y := T_y(U\otimes I_F),
\]
gives a $\QAC$ circuit $U_y$ of width $N+1$ and depth at most $d+1$. Writing $J_0\ket{\psi} = \ket{\psi}\otimes\ket{0}_F$ for the flag-appending isometry, we have $U_y J_0\ket{\psi} = (I-P_y)U\ket{\psi}\otimes\ket{0}_F + P_y U\ket{\psi}\otimes\ket{1}_F$, so the augmented acceptance operator $E_y = U_y^\ad(I\otimes\ket{1}\bra{1}_F)U_y$ satisfies
\begin{equation}
  J_0^\ad E_y J_0 = (P_y U)^\ad(P_y U) = U^\ad P_y U,
  \label{eq:flag-effect}
\end{equation}
and hence $\tr(E_y J_0\rho J_0^\ad) = \tr(P_y U\rho U^\ad)$ for every input state $\rho$.

When $X$ is an $m$-qubit designated register of $U$, keeping the same $X$ for $U_y$ and placing $F$ in the auxiliary register $A \otimes F$, any Hermitian approximant $\widetilde{B} = \sum_{\sigma\in\Pauli}\sigma\otimes\widetilde{B}_\sigma$ to $E_y$ compresses to
\begin{equation}
  B = J_0^\ad\widetilde{B} J_0
    = \sum_{\sigma\in\Pauli}\sigma\otimes
      \bigl((I_A\otimes\bra{0}_F)\widetilde{B}_\sigma(I_A\otimes\ket{0}_F)\bigr),
  \label{eq:flag-compression}
\end{equation}
which is Hermitian and satisfies $\deg_X(B) \le \deg_X(\widetilde{B})$ and $\norm{U^\ad P_y U - B} \le \norm{E_y - \widetilde{B}}$ by~\eqref{eq:flag-effect} and $\norm{J_0} = 1$.

\subsection{Pseudorandomness of Quantum State Designs}\label{sec:state-designs}

Write $\rho_\psi = \ket{\psi}\bra{\psi}$ and let $\mathcal{H}_m$ denote the Haar ensemble of $m$-qubit pure states. An ensemble $\mathcal{D}$ of $m$-qubit pure states is an $\varepsilon_{\mathrm{des}}$-approximate state $2$-design if its second-moment error satisfies
\[
  \norm{
    \mathbb{E}_{\psi\sim\mathcal{D}}\rho_\psi^{\otimes 2}
    - \mathbb{E}_{\phi\sim\mathcal{H}_m}\rho_\phi^{\otimes 2}
  }_1 \le \varepsilon_{\mathrm{des}},
\]
where $\norm{\cdot}_1$ denotes the trace norm; a design is \emph{exact} if $\varepsilon_{\mathrm{des}} = 0$.

\begin{theorem}[Pseudorandomness of State $2$-Designs]\label{th:state-designs}
Fix $d \in \mathbb{Z}_{\ge 1}$, $\kw \ge 1$, $\alpha \in (0,1)$, and $c > 0$. For all sufficiently large $m$, let $\mathcal{D}$ be an $m$-qubit state $2$-design with second-moment error at most $2^{-cm}$. Let $U$ be a single-output $\QAC$ circuit of depth at most $d$ and width $N \le m^{\kw}$. For any positive integer $t$ with $tm \le N$ and any fixed state $\omega$ on the remaining $N-tm$ qubits, define the acceptance probability
\[
  p(\psi) = \tr\bigl(U^\ad \Pout U\,(\rho_\psi^{\otimes t}\otimes\omega)\bigr).
\]
Then
\[
  \left|\mathbb{E}_{\psi\sim\mathcal{D}}p(\psi)
        - \mathbb{E}_{\phi\sim\mathcal{H}_m}p(\phi)\right| \le e^{-m^\alpha}.
\]
\end{theorem}

\begin{proof}
Write $E=U^\ad\Pout U$, $D_X=2^m$, and $\mu=I/D_X$. Fix $\alpha' \in (\alpha, 1)$ and set $\delta=(1-\alpha')2^{-d}/2$, $r=\lfloor m^{1-\delta}\rfloor$, and $\varepsilon=e^{-m^{\alpha'}}$. For any designated copy register $X$ (with all other qubits in $A$), since $m \le tm \le N \le m^{\kw}$, Theorem~\ref{th:approximation} with parameters $(\alpha', \eta = \delta)$ gives a Hermitian operator $B$ satisfying $\norm{E-B} \le \varepsilon$ and $\deg_X(B) \le r$.

For any density operator $\nu$ on $A$, define the Hermitian operator $B_\nu = \sum_{\sigma\in\Pauli} b_\sigma\sigma$ with real coefficients $b_\sigma = \tr(\nu B_\sigma)$, so $\tr((Y\otimes\nu)B) = \tr(YB_\nu)$ for every operator $Y$ on $X$. Taking $Y = \ket{v}\bra{v}$ over unit vectors $\ket{v}$ gives $\norm{B_\nu} \le \norm{B} \le 1 + \varepsilon$, and Pauli orthogonality yields $\sum_{\sigma\in\Pauli}|b_\sigma|^2 = D_X^{-1}\tr(B_\nu^2) \le \norm{B_\nu}^2 \le (1+\varepsilon)^2$. Since $\tr(\rho_\psi - \mu) = 0$, $\tr(\sigma\mu) = 0$ for $\sigma \ne I_X$, and $b_\sigma = 0$ for $\wt(\sigma) > r$, setting $S_r(\psi) = \sum_{1\le\wt(\sigma)\le r}|\tr(\sigma\rho_\psi)|^2$ and applying Cauchy--Schwarz together with $\norm{(\rho_\psi-\mu)\otimes\nu}_1 = \norm{\rho_\psi-\mu}_1 \le 2$ gives
\begin{equation}
  \left|\tr\bigl(E\,((\rho_\psi-\mu)\otimes\nu)\bigr)\right|
  \le 2\varepsilon+(1+\varepsilon)S_r(\psi)^{1/2}.
  \label{eq:design-one-copy}
\end{equation}

By two-copy Schur--Weyl duality, $\mathbb{E}_{\phi\sim\mathcal{H}_m}\rho_\phi^{\otimes 2}$ is the normalized symmetric-subspace projector $(I+F)/(D_X(D_X+1))$, where $F$ swaps the two copies. Since $\tr((\sigma\otimes\sigma)F) = \tr(\sigma^2) = D_X$ and $\tr(\sigma) = 0$ for every $\sigma \in \Pauli \setminus \{I_X\}$,
\[
  \mathbb{E}_{\phi\sim\mathcal{H}_m}|\tr(\sigma\rho_\phi)|^2
  = \frac{\tr(\sigma)^2+\tr(\sigma^2)}{D_X(D_X+1)}
  = \frac{1}{D_X+1},
\]
and $\norm{\sigma\otimes\sigma} = 1$ implies $\mathbb{E}_{\psi\sim\mathcal{D}}|\tr(\sigma\rho_\psi)|^2 \le 1/(D_X+1) + 2^{-cm}$. The number of non-identity Paulis of weight at most $r$ satisfies
\[
  M_r = \sum_{j=1}^r 3^j\binom{m}{j} \le (3m)^{r+1} = 2^{o(m)},
\]
so Jensen's inequality gives $\mathbb{E}_{\psi\sim\mathcal{E}}[S_r(\psi)^{1/2}] \le \bigl(M_r(1/(D_X+1)+2^{-cm})\bigr)^{1/2}$ for both $\mathcal{E} \in \{\mathcal{D}, \mathcal{H}_m\}$.

Setting $\tau_j(\psi) = \mu^{\otimes j}\otimes\rho_\psi^{\otimes(t-j)}\otimes\omega$ ($0 \le j \le t$) and $\tau_t := \tau_t(\psi) = \mu^{\otimes t}\otimes\omega$, applying~\eqref{eq:design-one-copy} to the $j$-th copy register in each difference $\tau_{j-1}(\psi) - \tau_j(\psi)$ yields
\[
  \left|p(\psi)-\tr(E\tau_t)\right|
  \le \sum_{j=1}^t \left|\tr\bigl(E(\tau_{j-1}(\psi)-\tau_j(\psi))\bigr)\right|
  \le t\bigl(2\varepsilon+(1+\varepsilon)S_r(\psi)^{1/2}\bigr).
\]
Averaging over $\mathcal{D}$ and $\mathcal{H}_m$ and comparing both to $\tr(E\tau_t)$ bounds their acceptance-probability difference by
\[
  4t\varepsilon+2t(1+\varepsilon)\bigl(M_r(1/(D_X+1)+2^{-cm})\bigr)^{1/2}
  = o(e^{-m^\alpha}) + 2^{-\Omega(m)} \le e^{-m^\alpha}
\]
for all sufficiently large $m$, using $t \le m^{\kw}$, $\alpha' > \alpha$, $D_X = 2^m$, and $t^2 M_r = 2^{o(m)}$.
\end{proof}

\begin{corollary}[Multi-Qubit Readout]\label{cor:state-designs-multi}
Fix $d \in \mathbb{Z}_{\ge 1}$, $\kw \ge 1$, $\alpha, \beta \in (0,1)$, and $c > 0$. In the setting of Theorem~\ref{th:state-designs}, measuring $1 \le s \le m^\beta$ designated output qubits of $U$ in the computational basis yields output distributions under $\mathcal{D}$ and $\mathcal{H}_m$ whose total variation distance is at most $e^{-m^\alpha}$ for all sufficiently large $m$.
\end{corollary}

\begin{proof}
Let $p_{\mathcal{E}}(y)$ denote the probability of outcome $y \in \{0,1\}^s$ under $\mathcal{E} \in \{\mathcal{D}, \mathcal{H}_m\}$. By~\eqref{eq:flag-effect}, $p_{\mathcal{E}}(y)$ equals the average acceptance probability of $U_y$ (depth at most $d+1$, width $N+1 \le m^{\kw+1}$ for $m \ge 2$) with auxiliary state $\omega\otimes\ket{0}\bra{0}_F$. Applying Theorem~\ref{th:state-designs} with $\alpha' \in (\max\{\beta,\alpha\}, 1)$ gives $|p_{\mathcal{D}}(y) - p_{\mathcal{H}_m}(y)| \le e^{-m^{\alpha'}}$ for each $y$, so
\[
  \frac{1}{2}\sum_{y\in\{0,1\}^s}|p_{\mathcal{D}}(y)-p_{\mathcal{H}_m}(y)|
  \le 2^{s-1}e^{-m^{\alpha'}}
  \le \exp\!\bigl(m^\beta\log 2-m^{\alpha'}\bigr)
  \le e^{-m^\alpha}
\]
for all sufficiently large $m$.
\end{proof}

\paragraph{Stabilizer testing.}
For a uniformly random $m$-qubit Clifford unitary $V$, compare $V\ket{0^m}$ and $V\ket{\chi}$, where $\ket{\chi}=\ket{T}\otimes\ket{0^{m-1}}$ and $\ket{T}=(\ket{0}+e^{\iu\pi/4}\ket{1})/\sqrt{2}$. Because the Clifford group is a unitary $3$-design~\cite{Zhu17}, both ensembles match the Haar second moment and are exact state $2$-designs, so Theorem~\ref{th:state-designs} and the triangle inequality bound their distinguishing advantage by $2e^{-m^\alpha}$. On the other hand, for any pure stabilizer state $\ket{s}$ with stabilizer group $\mathcal{S}$, tracing out qubits $2,\dots,m$ from $\ket{s}\bra{s} = 2^{-m}\sum_{P\in\mathcal{S}}P$ leaves only commuting Paulis on qubit~$1$—hence either $I/2$ or a Pauli eigenstate. Since $\ket{T}\bra{T} = \frac{1}{2}(I + (\mathrm{X}+\mathrm{Y})/\sqrt{2})$, we have
\[
  |\braket{s}{\chi}|^2
  \le \bra{s}(\ket{T}\bra{T}\otimes I)\ket{s}
  \le \frac{1+1/\sqrt{2}}{2}=\cos^2(\pi/8).
\]
Since $V^\ad$ permutes pure stabilizer states, every state $V\ket{\chi}$ has trace distance $\sqrt{1 - |\braket{s}{V\ket{\chi}}|^2} \ge \sin(\pi/8)$ from every pure stabilizer state $\ket{s}$. Consequently, the six-copy test of Gross, Nezami, and Walter~\cite[Theorem~3.3]{GNW21} accepts $V\ket{0^m}$ with probability $1$ and $V\ket{\chi}$ with probability at most $1 - \sin^2(\pi/8)/4$, giving a constant acceptance gap.

\subsection{Low-Energy State Preparation}\label{sec:low-energy}

Anshu, Breuckmann, and Nirkhe~\cite{ABN23} proved that preparing low-energy states of their NLTS Hamiltonians requires depth $\Omega(\log m)$ with bounded-fan-in gates, and Anshu et~al.~\cite[Corollary~6.8]{ADOY25} obtained a superlinear ancilla lower bound for $\QAC$ circuits preparing low-energy pure states to high fidelity.

Consider a CSS code family of block length $m$ encoding $k_m$ logical qubits, specified by binary check matrices $M_Z, M_X$ with $q_Z, q_X$ rows and $M_Z M_X^{\mathsf{T}} = 0$. Rows $z$ of $M_Z$ and $x$ of $M_X$ define checks $\mathrm{Z}^z = \bigotimes_i \mathrm{Z}^{z_i}$ and $\mathrm{X}^x = \bigotimes_i \mathrm{X}^{x_i}$, indexed altogether as $S_j$ ($Z$-checks first), with local Hamiltonian $\Ham_m = \sum_j (I - S_j)/2$. Writing $|v|$ for Hamming weight, the family has \emph{linear-distance low-syndrome clustering}~\cite[Property~1]{ABN23} if it has distance at least $\lambda m$ and $\min(q_Z, q_X) \ge h m$ for constants $\lambda, h > 0$, and there exist constants $c_1, c_2 > 0$ such that for all sufficiently small $\xi > 0$: every $x \in \mathbb{F}_2^m$ with $|M_Z x| \le \xi q_Z$ has reduced weight $d_X(x) := \min_{u\in\operatorname{row}(M_X)}|x+u|$ either at most $c_1\xi m$ or at least $c_2 m$, and similarly every $z \in \mathbb{F}_2^m$ with $|M_X z| \le \xi q_X$ has $d_Z(z) := \min_{v\in\operatorname{row}(M_Z)}|z+v|$ either at most $c_1\xi m$ or at least $c_2 m$.

The constant-rate, bounded-check-weight CSS codes of~\cite{ABN23} satisfy these properties. Indeed, for any CSS family of distance at least $\lambda m$ with $k_m > 0$, reduced row echelon form yields a basis of $\ker M_Z$ whose vectors each have weight at most $\operatorname{rank} M_Z + 1$; since $\dim(\ker M_Z/\operatorname{row}(M_X)) = k_m > 0$, at least one basis vector represents a nontrivial logical $\mathrm{X}$ operator of weight at least $\lambda m$, so $q_Z \ge \operatorname{rank} M_Z \ge \lambda m - 1$ and likewise $q_X \ge \lambda m - 1$ (giving $\min(q_Z,q_X) \ge (\lambda/2)m$ for $m \ge 2/\lambda$).

\begin{theorem}[Low-Energy Spectral Weight]\label{th:low-energy-qac}
Let $(\Ham_m)$ be the local Hamiltonian family of a CSS code family encoding $k_m$ logical qubits with linear-distance low-syndrome clustering. There exists a constant $e_0 > 0$ depending only on the code family such that the following holds. Fix $d \in \mathbb{Z}_{\ge 1}$, $\kw \ge 1$, and $\alpha \in (0,1)$. For all sufficiently large $m$ in the family, let $U$ be an $N$-qubit $\QAC$ circuit of depth at most $d$ with $N \le m^{\kw}$ and $m$ designated output qubits. If $b < k_m$ input qubits are initialized in an arbitrary joint state and all remaining input qubits in $\ket{0}$, then the reduced output state $\rho$ on the $m$ output qubits satisfies
\[
  \tr\bigl([\Ham_m \le e_0 m]\,\rho\bigr) \le 2e^{-m^\alpha}.
\]
\end{theorem}

\begin{proof}
Let $X$ be the $m$-qubit designated output register and $A$ the remaining $N-m$ output qubits. For initial state $\rho_{\mathrm{in}}$ on the $b$ arbitrary input qubits, the full output state $\sigma_{\mathrm{out}} = U(\rho_{\mathrm{in}}\otimes\ket{0^{N-b}}\bra{0^{N-b}})U^\ad$ satisfies $P_{\mathrm{prep}}\sigma_{\mathrm{out}}P_{\mathrm{prep}} = \sigma_{\mathrm{out}}$ for the rank-$2^b$ projector
\[
  P_{\mathrm{prep}} = U\bigl(I_{2^b} \otimes \ket{0^{N-b}}\bra{0^{N-b}}\bigr)U^\ad.
\]
Reversing $U$ gives a depth-$d$ $\QAC$ circuit for $U^\ad$. Applying the flag construction to test the all-zero pattern on the $N-b \ge m - k_m + 1 \ge 1$ clean qubits of $U^\ad$ yields a circuit of depth at most $d+1$ and width $N+1 \le m^{\kw+1}$ (for $m \ge 2$) whose acceptance operator compresses to $P_{\mathrm{prep}}$ via~\eqref{eq:flag-effect}. Applying Theorem~\ref{th:approximation} with $\eta = (1-\alpha)2^{-(d+2)}$ and compressing via~\eqref{eq:flag-compression} yields a Hermitian operator $B$ on $X \otimes A$ satisfying
\[
  \norm{P_{\mathrm{prep}}-B}\le \varepsilon_m=e^{-m^\alpha},
  \qquad
  \deg_X(B)\le r_m=\bigl\lfloor m^{1-(1-\alpha)2^{-(d+2)}}\bigr\rfloor=o(m).
\]

\paragraph{Representatives of low-energy syndromes.}
Fix $\xi > 0$ in the clustering regime with $c_1\xi < c_2/2$ and $4c_1\xi < \lambda/2$, and set $e_0 = \xi h/3$. For each syndrome $s = (s_Z, s_X)$, let $\mathcal{C}_s$ be the joint eigenspace where each $S_i$ acts as $(-1)^{s_i}$, so $\Ham_m = |s|$ on $\mathcal{C}_s$. On any nonzero $\mathcal{C}_s$, the eigenvalues of $m - k_m$ independent stabilizer generators $G_a$, say $\varepsilon_a(s) \in \{\pm 1\}$, determine those of all remaining checks, and every nonempty product of independent $G_a$'s is a traceless Pauli; hence
\[
  \dim\mathcal{C}_s
  = \tr\!\left(\prod_{a=1}^{m-k_m}\frac{I+\varepsilon_a(s)G_a}{2}\right)
  = 2^{-(m-k_m)}2^m
  = 2^{k_m}.
\]

Up to phase, $P = \mathrm{X}^x\mathrm{Z}^z$ has syndrome $\operatorname{syn}(P) = (M_Z x, M_X z)$ and maps $\mathcal{C}_s$ onto $\mathcal{C}_{s+\operatorname{syn}(P)}$. Form an undirected graph on $\Xi = \{s : \mathcal{C}_s \ne 0,\ |s| \le e_0 m\}$ by joining $s, t \in \Xi$ whenever some Pauli of weight at most $r_m$ maps $\mathcal{C}_s$ onto $\mathcal{C}_t$, and fix a root $s_0$ in each connected component. Along any path $s_0, s_1, \dots, s_\ell = s$ with edge-witnessing Paulis $P_j$ of weight at most $r_m$, write $P_j\cdots P_1 = \mathrm{X}^{x_j}\mathrm{Z}^{z_j}$ up to phase; its syndrome is $s_j + s_0$, so
\[
  |M_Z x_j|,\ |M_X z_j| \le |s_j|+|s_0| \le 2e_0m \le \xi\min(q_Z,q_X).
\]
Since $d_X$ and $d_Z$ are $1$-Lipschitz in Hamming distance, both start at $0$ and change by at most $r_m$ per step. For sufficiently large $m$, $c_1\xi m + r_m < c_2 m$, so clustering keeps $d_X(x_j), d_Z(z_j) \le c_1\xi m$ inductively at every step. Choosing $u \in \operatorname{row}(M_X)$ and $v \in \operatorname{row}(M_Z)$ attaining $d_X(x_\ell)$ and $d_Z(z_\ell)$ for a fixed path to $s$, the Pauli $V_s = \mathrm{X}^{x_\ell+u}\mathrm{Z}^{z_\ell+v}$ has weight at most $2c_1\xi m$ and syndrome $(M_Z x_\ell, M_X z_\ell) = s + s_0$, hence maps $\mathcal{C}_{s_0}$ onto $\mathcal{C}_s$.

\paragraph{Logical multiplicity and rank comparison.}
Transport a fixed orthonormal basis of each root space $\mathcal{C}_{s_0}$ to $\mathcal{C}_s$ via $V_s$. For any Pauli $P$ of weight at most $r_m$ and $s, t \in \Xi$, the block of $P$ from $\mathcal{C}_s$ to $\mathcal{C}_t$ vanishes unless $t = s + \operatorname{syn}(P)$; in that case $s$ and $t$ share the same root $s_0$, and $V_t^\ad P V_s$ preserves $\mathcal{C}_{s_0}$ and has weight at most $4c_1\xi m + r_m < \lambda m$ (for large $m$), so by code distance it is a stabilizer up to phase and acts as a scalar multiple of $I_{2^{k_m}}$ on $\mathcal{C}_{s_0}$. Setting $Q = [\Ham_m \le e_0 m]\otimes I_A$ and collecting these scalars into matrices $T_\sigma$ on the syndrome space yields, in the transported coordinates on $\operatorname{im}(Q)$,
\[
  QBQ \simeq I_{2^{k_m}}\otimes
       \left(\sum_{\substack{\sigma\in\Pauli\\\wt(\sigma)\le r_m}} T_\sigma\otimes B_\sigma\right),
\]
where the parenthesized operator is Hermitian since $QBQ$ is; every eigenspace of $QBQ$ in $\operatorname{im}(Q)$ therefore has dimension at least $2^{k_m}$.

Finally, $\norm{QP_{\mathrm{prep}}Q - QBQ} \le \varepsilon_m$ while $\operatorname{rank}(QP_{\mathrm{prep}}Q) \le 2^b < 2^{k_m}$. As in~\cite{WHLL26}, by rank--nullity every eigenspace of $QBQ$ in $\operatorname{im}(Q)$ contains a unit vector $\ket{v}$ with $QP_{\mathrm{prep}}Q\ket{v} = 0$, so its eigenvalue is $\bra{v}QBQ\ket{v} \le \varepsilon_m$. Thus $QBQ \preceq \varepsilon_m Q$ and $0 \preceq QP_{\mathrm{prep}}Q \preceq QBQ + \varepsilon_m Q \preceq 2\varepsilon_m Q$. Since $\norm{P_{\mathrm{prep}}QP_{\mathrm{prep}}} = \norm{QP_{\mathrm{prep}}}^2 = \norm{QP_{\mathrm{prep}}Q}$ and $\rho = \tr_A(\sigma_{\mathrm{out}})$, we conclude that
\[
  \tr\bigl([\Ham_m\le e_0m]\rho\bigr)
  = \tr(Q\sigma_{\mathrm{out}})
  = \tr(P_{\mathrm{prep}}QP_{\mathrm{prep}}\sigma_{\mathrm{out}})
  \le \norm{QP_{\mathrm{prep}}Q}
  \le 2\varepsilon_m. \qedhere
\]
\end{proof}

Since $\Ham_m \succeq e_0 m[\Ham_m > e_0 m]$, Theorem~\ref{th:low-energy-qac} implies $\tr(\Ham_m\rho) \ge e_0 m(1 - 2e^{-m^\alpha})$ whenever $b < k_m$; in particular, reaching energy at most $e_0 m$ with inverse-polynomial probability requires at least $k_m$ arbitrary input qubits.

\section{Concluding Remarks and Open Problems}\label{sec:open}

A natural question left open by our separation $\AC \not\subseteq \QAdv$ is whether $\QAC$ circuits can compute the indexing function in the single-copy, single-output model, a problem also highlighted by Grier, Morris, and Wu~\cite[Section~1.2]{GMW26}. For $k \ge 1$ and $M = 2^k$, define $\mathrm{INDEX}_M \colon \{0,1\}^k \times \{0,1\}^M \to \{0,1\}$ by $\mathrm{INDEX}_M(i,x) = x_i$, where $x = (x_j)_{j \in \{0,1\}^k}$. On $n = M + k$ input bits, the family $\mathrm{INDEX} = (\mathrm{INDEX}_{2^k})_{k \ge 1}$ lies in $\AC$, as a depth-$2$ DNF with $M$ terms of width $k+1$ computes it using $O(M\log M)$ literal occurrences.

Existing degree and Fourier methods do not resolve the complexity of $\mathrm{INDEX}$. Writing $\ell_0(u)=1-u$ and $\ell_1(u)=u$, the real polynomial $\sum_{j\in\{0,1\}^k} x_j\prod_{a=1}^k\ell_{j_a}(i_a)$ agrees with $\mathrm{INDEX}_M(i,x)$ on $\{0,1\}^{k+M}$, giving exact degree at most $k+1$ and Fourier support on levels at most $k+1$. Conversely, introducing a fresh bit $b$ and substituting $x_j = b \oplus \mathrm{PARITY}_k(j) \in \{b, 1-b\}$ yields $\mathrm{INDEX}_M(i,x) = x_i = \mathrm{PARITY}_{k+1}(i,b)$ without increasing polynomial degree: if $q(i,b)$ approximates $\mathrm{PARITY}_{k+1}(i,b)$ within $\varepsilon \in (0, 1/2)$, then $\chi(i,b) = (-1)^{b+\sum_a i_a}$ satisfies $\mathbb{E}_{i,b}[\chi(i,b)(1 - 2q(i,b))] \ge 1 - 2\varepsilon > 0$ under the uniform distribution, whereas every monomial of degree at most $k$ omits at least one bit and correlates to zero with $\chi$, forcing $\deg q \ge k+1$ and hence $\ade_\varepsilon(\mathrm{INDEX}_M) = k+1$.

Because $k+1 = O(\log n)$, this degree is compatible with both Corollary~\ref{cor:degree} and the uniform-input Fourier-tail bound of~\cite[Corollary~9]{OAI26}. Likewise, restricting $x_j = \mathrm{PARITY}_k(j)$ yields $\mathrm{PARITY}_k$ on the $m = k$ free address bits, but the restricted circuit retains all $M = 2^k$ fixed data qubits in the auxiliary register, so its width $N \ge 2^k + k$ is exponential in $m$ rather than $m^{O(1)}$ as required by Theorem~\ref{th:approximation}. Determining whether $\mathrm{INDEX}$ belongs to $\BQ$ or $\QAdv$—and, for each fixed depth $d \ge 1$, the minimum width needed to compute $\mathrm{INDEX}_M$ with bounded error—thus appears to require techniques beyond polynomial approximation and polynomial-width input restrictions.

\section*{AI Disclosure}

The proofs in this paper took shape through a series of discussions with
OpenAI's ChatGPT in which we explored and refined different approaches.
We used both ChatGPT and Google's Gemini to help draft and revise the
manuscript. We take responsibility for the mathematical claims, proofs,
and final text.

\end{document}